\documentclass{article}

\usepackage{arxiv}

\usepackage[utf8]{inputenc} % allow utf-8 input
\usepackage[T1]{fontenc}    % use 8-bit T1 fonts
\usepackage{hyperref}       % hyperlinks
\usepackage{url}            % simple URL typesetting
\usepackage{booktabs}       % professional-quality tables
\usepackage{amsfonts}       % blackboard math symbols
\usepackage{nicefrac}       % compact symbols for 1/2, etc.
\usepackage{microtype}      % microtypography
\usepackage{lipsum}
\usepackage{amsmath,amssymb}
\usepackage{graphicx}
\usepackage{cite}
\usepackage{hyper ref}
\usepackage{amssymb}
\usepackage{threeparttable}

\title{Solar Radio Spectro-polarimetry (50 - 500 MHz) : Design and Development of Cross-Polarized Log-Periodic Dipole antenna and configuration of receiver system}

\author{
  Anshu Kumari\thanks{Anshu Kumari was with the Indian Institute of Astrophysics, Bangalore earlier. And is presently with the University of Helsinki Finland.)} \\
  Department of Physics\\
 University of Helsinki \\
 P.O. Box 64, FI-00014 Helsinki, Finland \\
  \texttt{anshu.kumari@helsinki.fi} \\
   \And
 G. V. S. Gireesh \\
  Indian Institute of Astrophysics\\
  Koromangala II block \\
  Bangalore, India, 560034\\
  \texttt{gireesh@iiap.res.in} \\
  
     \And
 C. Kathiravan \\
  Indian Institute of Astrophysics\\
  Koromangala II block \\
  Bangalore, India, 560034\\
  \texttt{kathir@iiap.res.in} \\
  
     \And
 V. Mugundhan\thanks{V. Mugundhan was with the Indian Institute of Astrophysics, Bangalore earlier. And is presently with the Raman Research Institute, Bangalore.)} \\
  Raman Research Institute\\
  C. V. Raman Avenue, Sadashivanagar \\
  Bangalore, India, 560080\\
  \texttt{mugundhan@rri.res.in} \\
  
       \And
Indrajit V. Barve \\
  Indian Institute of Astrophysics\\
  Koromangala II block \\
  Bangalore, India, 560034\\
  \texttt{indrajit@iiap.res.in} \\

}

\begin{document}
\maketitle

\begin{abstract}
A radio spectro-polarimeter was developed at the Gauribidanur radio observatory (Longitude : $77^\circ 27' 07''$; Latitude : $13^\circ 36' 12''$) to study the characteristics of the polarized radio waves that are emitted by the impetuous solar corona in the 50 - 500 MHz frequency range. The instrument has three major components : a Cross-polarized Log-Periodic Dipole Antenna (CLPDA), an analog receiver, and a digital receiver (spectrum analyzer). This article elaborates the design and developmental aspects of the CLPDA, its characteristics and briefs about the configurations of the analog and digital receivers, setting up of the spectro-polarimeter, stage-wise tests performed to characterize it, etc. To demonstrate the instrumental capability, the estimation of the solar coronal magnetic field strength (B {\it Vs} heliocentric height), using the spectral data obtained with it, is exemplified.

Throughout the above band, the CLPDA has a gain, return loss and polarization cross-talk of $\approx$ 6.6 dBi, $\lesssim$ -10 dB, and $\lesssim$ -27 dB, respectively. 
The design constraints, the procedure to tune its impedance and to minimize its dimension, etc. are elaborated. The analog receiver has a noise figure of $\approx 3$ dB and a receiver-noise-temperature ($T_{rcvr}$) of about 290 K. The digital receiver can sweep and cover the above bandwidth in 4 ms (instantaneous bandwidth of $\approx$ 1.1 MHz). The spectral data acquired for ten successive sweeps are integrated (for 100 $\mu s$) and averaged onboard. The above parameters give a receiver-flux-density ($S_{rcvr}$) of $\approx 5.3 \times 10^3 $, and $\approx 5.3 \times 10^5$ Jy at 50 and 500 MHz, respectively. The observed spectral data shows a Signal-to-Noise Ratio and Dynamic range of about 30 dB and 40 dB, respectively, at 50 MHz. The average polarization isolation / cross-talk of the CLPD varies from -30 dB to -24 dB over an azimuthal angle of $\pm 45^\circ$ with respect to the reference position angle ($0^\circ$). The average degree of circular polarization (DCP) is $\approx 100\%$ at the reference position and found to decrease gradually and reaches $\approx 80\%$ at an azimuthal angle of $\pm 45^\circ$. The variation of the cross-talk and DCP as a function of azimuthal angle were used to have a one to one mapping in order to establish an association between cross-talk and DCP; the latter gives an uncertainty of $\approx 0.2, 2,$ and $20 \%$ in DCP for -30, -20 and -10 dB cross-talk,  respectively. The Stokes-I and Stokes-V spectrum of the type-V burst observed on March 30, 2018 with the SP was used to determine the associated magnetic field strength (B) as a function of heliocentric height. It was found that $B(r) \approx 16.8\pm 0.5 \,r^{-3.3}$ G.
\end{abstract}

% keywords can be removed

\section{Background}
\label{sec:background}
The performance of space borne technological systems depends on the weather conditions (called space weather) that prevail in the Earth's geospace \cite{mclean1985solar}. 
Space weather (SW) could be disastrous \cite{lin2000new} at times due to transient activities such as flares, coronal mass ejections (CMEs), etc. that take place in the outer solar atmosphere. Identifying the precursors of such events would therefore become essential to forecast SW reliably to safeguard the space borne systems. In the latter context, different types of radio outbursts (type-II, III, V, etc.) were recognized to be some of the precursors \cite{Cane2002type3} of above transients, especially in the low frequency radio regime. Observational studies show that the onset of aforementioned radio outbursts is predominantly decided by the strength, configuration, and spatio-temporal evolution of the associated solar active-region cum ambient magnetic field system (\cite{Allen47,Gopal1987type3}). Furthermore, the configuration of the latter plays an important role in deciding the polarization state and strength of the associated radio outbursts \cite{Allen47}. Therefore, one may utilize the observed polarization signatures, by virtue of unprecedented time and frequency resolution observations, to improve the prediction of SW forecasts. Since the aforesaid transient events were found to originate in the  low and middle corona (from where radio waves of 50 - 500 MHz emanate), designing a broad-band spectro-polarimeter to  observe this portion of coronal region would be advantageous. Therefore, the outcome of this exercise such as the estimates of density, temperature, magnetic field strength, etc. of the coronal region that are associated with above transients \cite{Allen47}, are expected to be helpful for SW forecasts and to study various aspects of transients, in general.

The technical content of this article is divided into five major sections to cover in detail, the design and developmental aspects of various components that constitute the spectro-polarimeter (SP) system and its usefulness in observing solar coronal transients: First, The design and fabrication of a broad-band Cross-Polarized Log-Periodic Dipole Antenna (CLPDA) is dealt elaborately (Section \ref{sec:DD_CLPDA}). Second, The field pattern (E \& H) and polarization cross-talk measurements of the CLPDA are discussed (Section \ref{sec:fab_CLPDA}). Third, The setting up of the SP by combining the CLPDA, an analog receiver and a digital receiver (Spectrum Analyzer), and the characterization of the system as a whole are explained (Section \ref{sec:SP_setup}). Fourth, The preliminary observations carried out and the results obtained using them subsequently, are presented to demonstrate the observing capability of the SP (Section \ref{sec:obs}). Fifth, the conclusions and future prospects are briefed (Section \ref{sec:vista}).

\section{Design and development of CLPDA}
\label{sec:DD_CLPDA}

The CLPDA, to outline simply, is a combination of two identical Log-Periodic Dipole Antennas (LPDAs); both LPDAs share a common vertical axis, however, the orientation of dipoles of one of the antennas is orthogonal \cite{Sasi2013} to the other; the vertical axis is a fictitious equi-divider line drawn in between the transmission lines of a LPDA. Obviously, to construct a CLPDA, two identical LPDAs are required at hand a priori.

\subsection{Introduction to LPDA}
\label{subsec:LPDA}
By definition, a LPDA is a coplanar array of dipoles; it has unequal length and unequally spaced parallel and linear dipoles that are fed alternatively ($180^\circ$ phase difference between adjacent dipoles) by a parallel transmission line with a desired characteristic impedance \cite{isbell1960log}. 
The electrical characteristics of it vary periodically with the logarithm of frequency (\cite{duhamel1966broadband,isbell1960log}), and hence the name LPDA.

\begin{figure}
\centering
\includegraphics[width=0.5\textwidth]{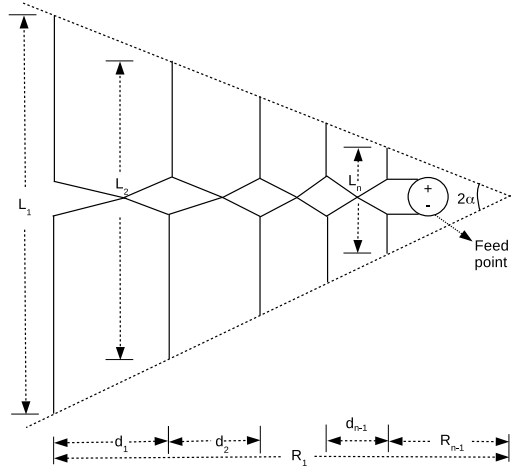}
\caption{Schematic diagram of a log periodic dipole antenna (not to scale).}
\label{fig:LPDA_schematic}
\end{figure}

\begin{figure}[!t]
\centering
\centerline{\includegraphics[width=0.5\textwidth]{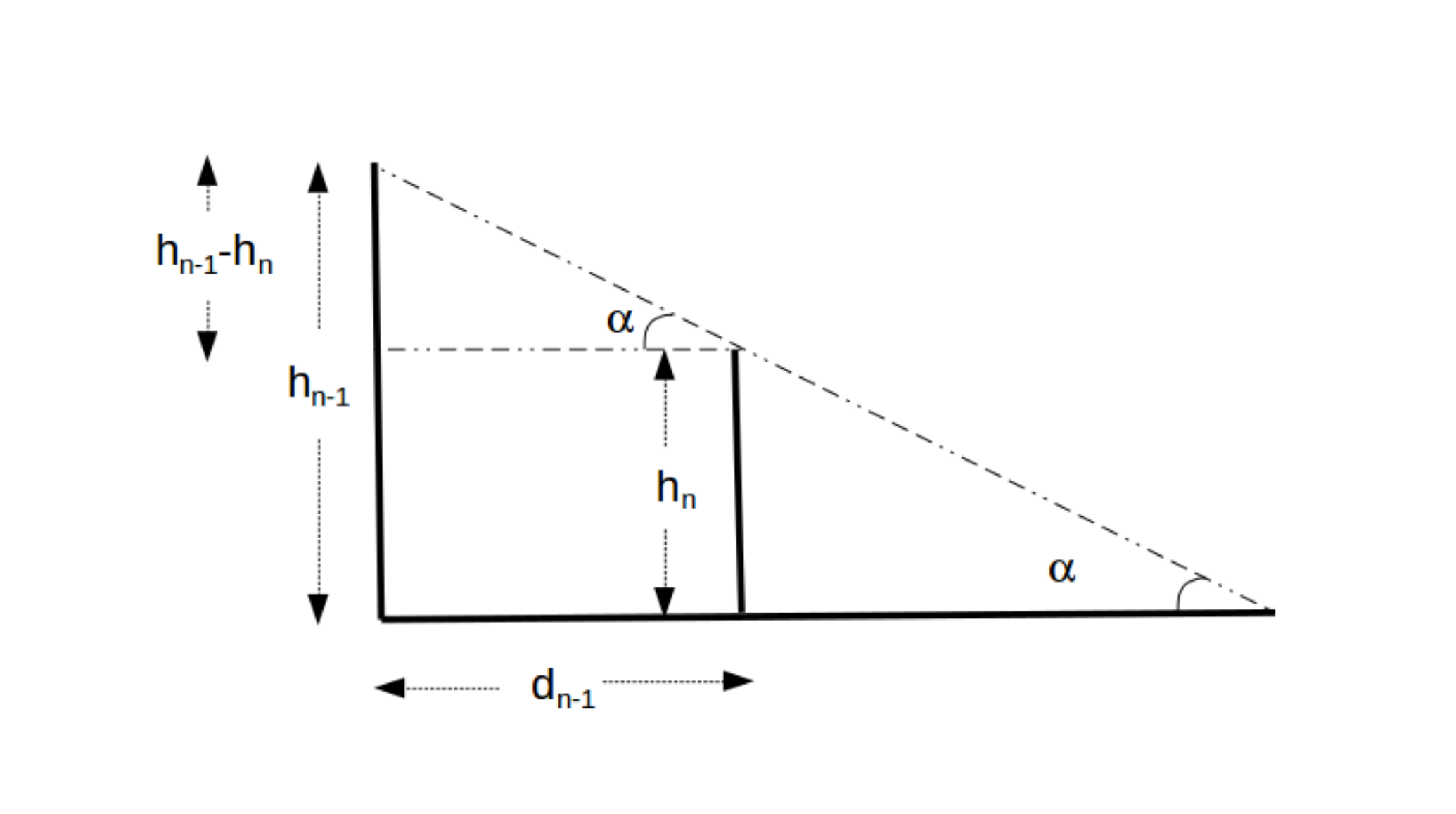}}
\caption{Schematic diagram showing the geometrical relationship between length of one-half of adjacent dipole arms (h), half-apex angle ($\alpha$) and inter-dipole spacing (d), of the LPDA. \label{fig:LPDA_geometry}}
\end{figure}

Empirical relationships suggested by Carrel \cite{carrel1961analysis} are usually followed to design a LPDA; refer Fig. \ref{fig:LPDA_schematic} for the schematic. 
The lowest and highest operating-frequencies, $f_1$ and $f_n$, respectively, are related to each other such that $f_1=\tau^{n-1}f_n$, where $\tau$ is the geometric constant, one of the design parameters. 
The relationship between length of adjacent dipole arms, $L_n$ and $L_{n-1}$, the spacing between them, $d_n$ and their distances from the apex, $R_n$ and $R_{n-1}$, respectively of a LPDA is given in Eqn. \ref{eq:LPDA_ratio}. 
\begin{equation}
\label{eq:LPDA_ratio}
\tau=\frac{L_n}{L_{n-1}}=\frac{R_n}{R_{n-1}}=\frac{d_{n-1}}{d_{n-2}}
\end{equation}
The other two design parameters are the spacing factor ($\sigma$), and the half apex angle ($\alpha$). By fixing any two design parameters, the third one can be determined using Eqn. \ref{eq:LPDA_alpha_sigma_tau}.
\begin{equation}
\label{eq:LPDA_alpha_sigma_tau}
\sigma=\frac{1-\tau}{4\tan\alpha}
\end{equation}
The geometrical relationship between the length of one-half of the dipole, half-apex angle and inter-dipole spacing of a LPDA is shown in Fig. \ref{fig:LPDA_geometry}. 
The length of the longest dipole is equal to half of the maximum wavelength ($\lambda_{max}$) of operation.

\subsection{New LPDA : Our design constraints and specifications}
\label{subsec:Dsgn_const_spec}
\begin{figure}[!t]
\centering
\centerline{\includegraphics[width=0.5\textwidth]{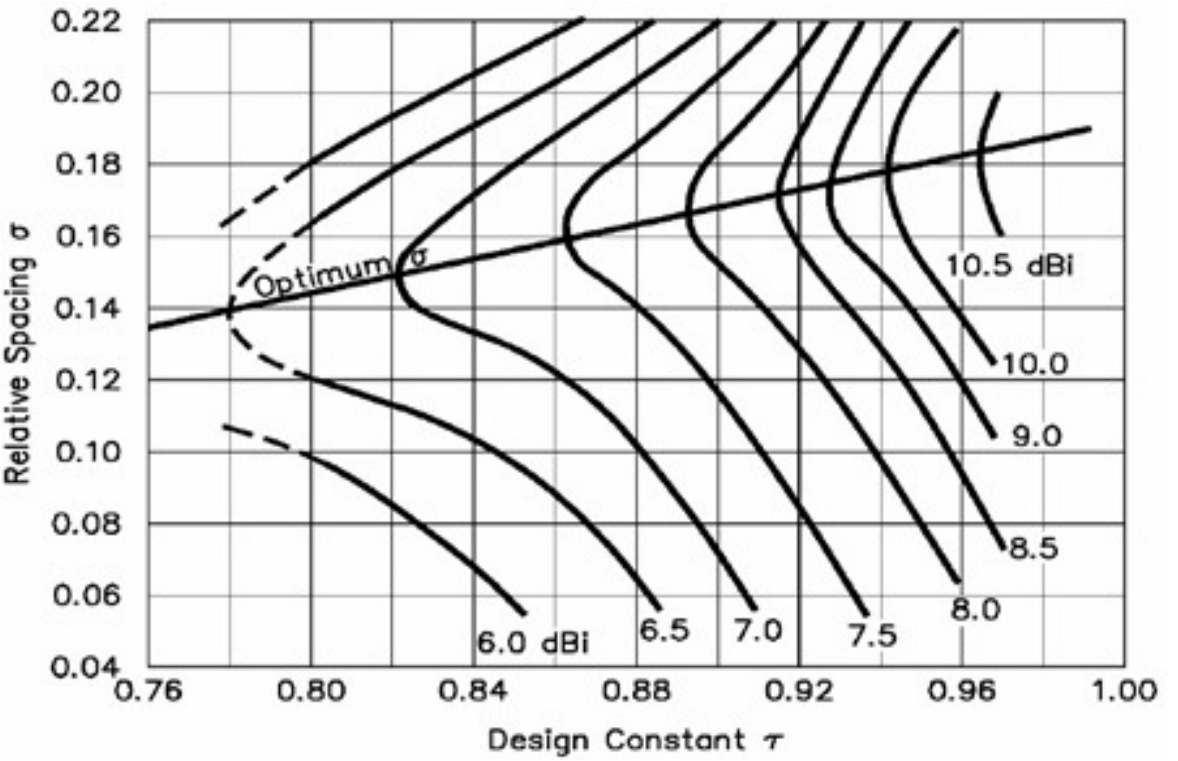}}
\caption{Plot of LPDA directional gain as a function of design parameters $\tau$ and $\sigma$ (reproduced from \cite{carrel1961analysis}) \label{fig:carrel_lpd_plot}}
\end{figure}

Having studied the performance of the LPDA extensively, Carrel\cite{carrel1961analysis} summarized its Directional Gain (G) as a function of $\tau$ and $\sigma$ (Fig. \ref{fig:carrel_lpd_plot}).
Since we decided to observe the celestial sources for a reasonably long duration, a moderate gain of around 6.4 dBi\footnote{Antenna gain with respect to an isotropic antenna.} was chosen for our prototype LPDA. Further, we wanted to fix an appropriate $\tau$ and $\sigma$ pair to design it; with reference to Fig. \ref{fig:carrel_lpd_plot}, we could find a range of $\tau$ and $\sigma$ values for a particular gain of the LPDA. In order to select a suitable pair, we varied $\tau$ and $\sigma$ between their lower and upper bounds. This exercise gave us an important input : The dimension of the LPDA grows as $\tau$ and $\sigma$ values are increased. Calculations showed that the dimension of our LPDA could vary from 1.5 m to 5.5 m. In order to minimize the difficulties in handling the LPDA, to optimize the mechanical requirements for mounting it to a rotor for tracking the Sun, and since the length of the longest dipole corresponding to the low-frequency cut-off was about 3 m, it was decided to fix the largest dimension of the antenna to be around 3 m; this constrained  the values of $\tau$ and $\sigma$ to be equal to 0.86 and 0.07, respectively. The total number of dipoles to cover the 50 - 500 MHz range (i.e, the operating bandwidth; shall be abbreviated as OB hereinafter) was determined using Eqn. \ref{eq:LPDA_N}; here $\beta$ is the bandwidth ratio.
\begin{equation}
\label{eq:LPDA_N}
{N=\frac{{ln}(\beta)}{{ln}(1/\tau)}}
\end{equation}
Substituting 10 for $\beta$ and 0.86 for $\tau$, we got $N=15$ dipoles for the LPDA; but since few more dipoles are usually added (conservative design) to realize the OB of the LPDA, we decided to prepare our prototype with 17 dipoles. The length of the dipoles, inter-dipole spacing, etc. used are listed in Table \ref{tab:LPDA_C1}.
\begin{table}
\caption{Design specification of the prototype LPDA  (50\,--\,500 MHz)}
\begin{center}
\begin{tabular}{ccccc}
\hline \\
\textbf{S. No.} & \textbf{Distance from} & \textbf{Inter-dipole} & \textbf{Half-Dipole} & \textbf{Frequency} \\
\textbf{} & \textbf{Ref. Pos.} & \textbf{spacing} & \textbf{Length ($\lambda / 4$)} & \textbf{($f=c/\lambda$)} \\
\textbf & \textbf{(cm)} & \textbf{(cm)} & \textbf{(cm)} & \textbf{(MHz)} \\ \hline
\\
1. & 0.0 & 0.0 & 13.5 & 555.6 \\ 
2. & 4.0 & 4.0 & 15.5 & 483.9 \\ 
3. & 9.0 & 5.0 & 18.0 & 416.7 \\ 
4. & 15.0 & 6.0 & 21.0 & 357.1 \\ 
5. & 22.0 & 7.0 & 24.5 & 306.1 \\ 
6. & 30.0 & 8.0 & 28.5 & 263.2 \\ 
7. & 39.0 & 9.0 & 33.0 & 227.3 \\ 
8. & 50.0 & 11.0 & 38.5 & 194.8 \\
9. & 63.0 & 13.0 & 45.0 & 166.7 \\ 
10. & 77.0 & 14.0 & 52.0 & 144.2 \\ 
11. & 94.0 & 17.0 & 60.5 & 124.0 \\ 
12. & 114.0 & 20.0 & 70.5 & 106.4 \\ 
13. & 137.0 & 23.0 & 82.0 & 91.5 \\ 
14. & 164.0 & 27.0 & 95.5 & 78.5 \\ 
15. & 195.0 & 31.0 & 111.0 & 67.6 \\ 
16. & 231.0 & 36.0 & 129.0 & 58.1 \\ 
17. & 273.0 & 42.0 & 150.0 & 50.0 \\ 
\hline
\end{tabular}
\end{center}
\label{tab:LPDA_C1}
\end{table}
Two square tubes (called booms), each having 25 mm $\times$ 25 mm sides, 4 mm wall thickness and 3 m length, were used as transmission lines; they were tied apart by 20 mm uniform inter-boom spacing with the help of insulators (called spacers) fixed at several locations along the boom. Cylindrical tubes of 13 mm diameter were used as dipoles and were fitted to the transmission lines as shown in Fig. \ref{fig:LPDA_schematic}. A feed-connector was fitted on the top side of the booms and just above the location wherein the shortest dipole (that would respond to the highest frequency) was fixed. The bottom side of the booms, close to the longest dipole was fixed, were shorted. A 3.2 m length RG-58 coax cable running through the grounded-boom was used as the feed cable to tap the output signal received by the LPDA. 

\subsection{The Impedance of the LPDA}
\label{subsec:Z_LPDA}
\begin{figure}[!t]
\centering
\centerline{\includegraphics[width=0.8\textwidth]{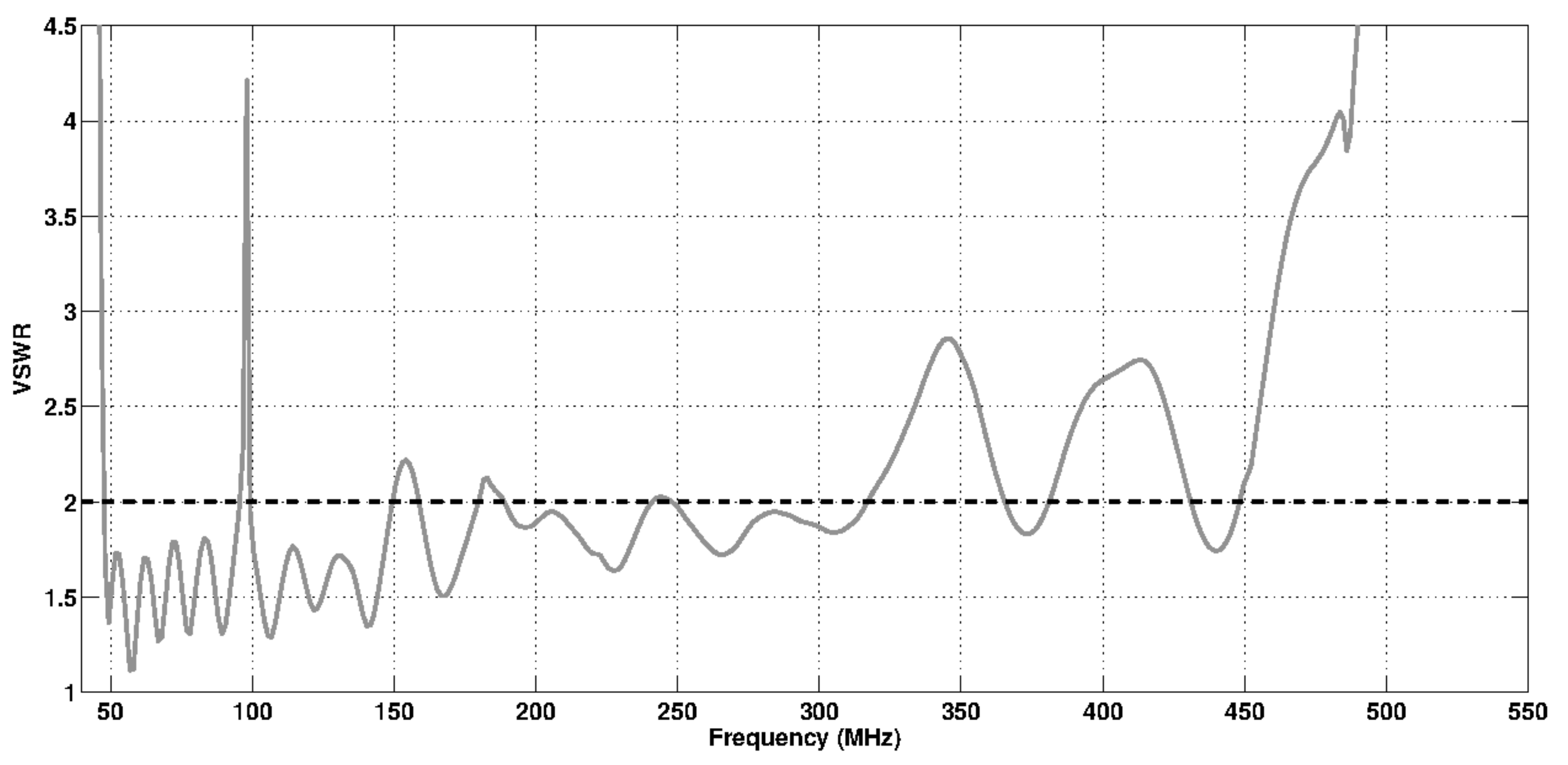}}
\caption{VSWR of the prototype LPDA as a function of frequency; the horizontal line drawn at VSWR equal to 2.0 shows a power transmission of $\approx 90\%$. The design specifications of the LPDA are given in Table \ref{tab:LPDA_C1}.\label{fig:LPDA_VSWR_1}}
\end{figure}

After the fabrication of the LPDA, its VSWR (Voltage Standing Wave Ratio), an indirect measure of impedance, was measured using a vector network analyzer; Fig. \ref{fig:LPDA_VSWR_1} shows the values as a function of frequency; it is clear that the values, over more than half of the OB, are above 2.0, the reference value corresponding to a power transmission of about $\approx 90\%$ between the source and the load.
To understand the observed trend, we compared the results of the experiments that were carried out to study the performance of LPDAs by varying the OB and design parameters : First, The impedance of the LPDA begins to approach the characteristic impedance ($Z_0=50\,\Omega$) throughout its OB when $\tau$ and $\sigma$ values are close to or greater than those of the optimal design curve shown in Fig. \ref{fig:carrel_lpd_plot}. Second, the VSWR spectrum becomes flat within the intended bandwidth if LPDAs are designed for slightly larger intended bandwidth. Third, The large value of $\tau$ and $\sigma$ increases the dimension of the LPDA as mentioned earlier. From Fig. \ref{fig:carrel_lpd_plot}, it can be seen that our selected parameter pair lies close to the lower edge of the 6.4 dBi gain-curve. Therefore, the impedance of the prototype LPDA may not lie close to $Z_0$, and consequently the weak cosmic signals cannot be received by it effectively throughout the OB. 

In order to improve the reception efficiency, it was decided to bring down the VSWR values below 2.0 over the entire OB. The procedure followed to realize that is elaborated in the following subsection.

\subsection{Fine-tuning the Impedance}
\label{subsec:finetune}
\begin{figure}[!t]
\centering
\centerline{\includegraphics[width=0.5\textwidth]{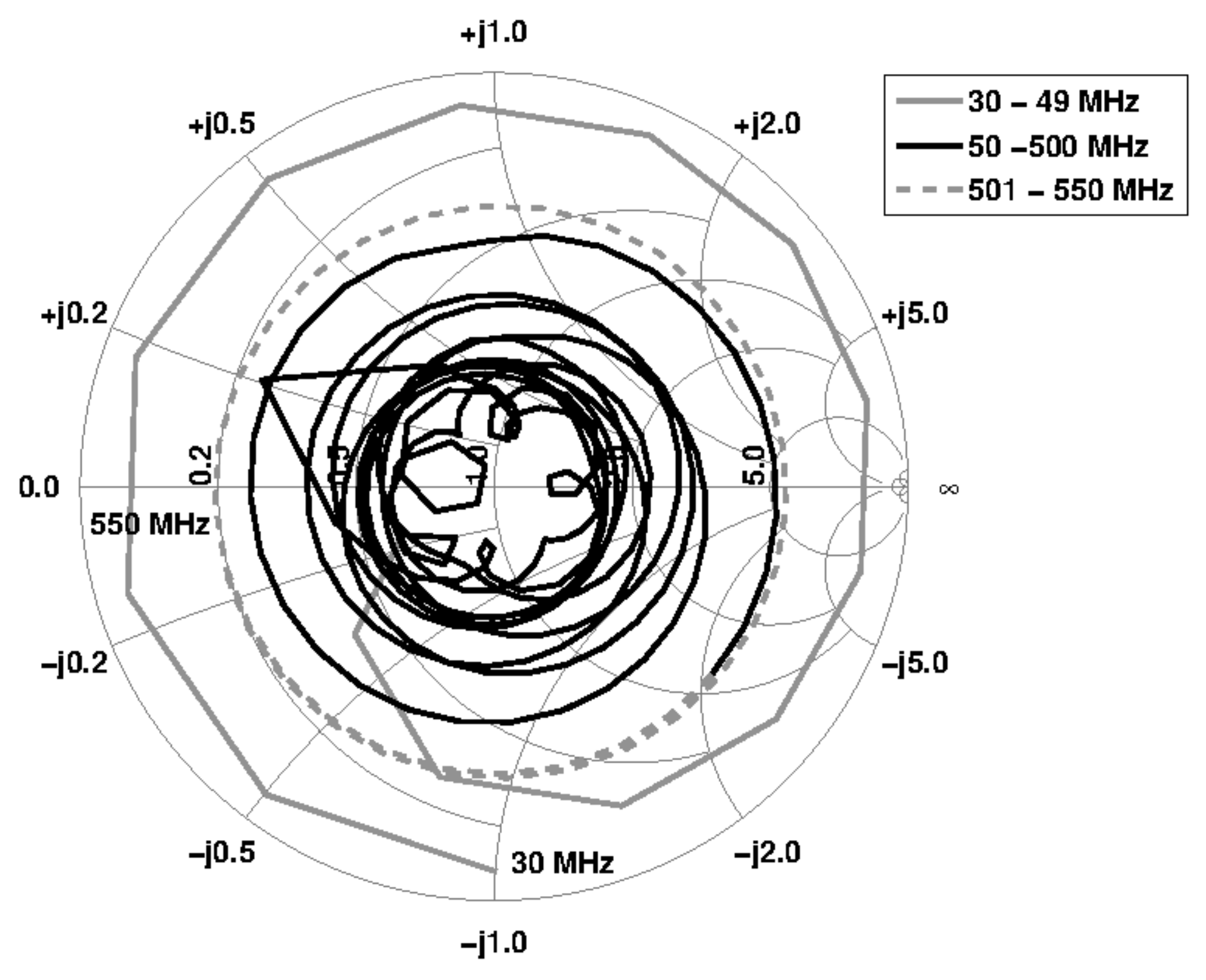}}
\caption{Smith chart / Impedance chart corresponding to Fig. \ref{fig:LPDA_VSWR_1}. \label{fig:LPDA_Smith_1}}
\end{figure}

To comprehend the variation of impedance of the prototype LPDA, with respect to frequency, the Smith chart was plotted, which is shown in Fig. \ref{fig:LPDA_Smith_1}; the values in the frequency range 30 - 49 MHz, 50 - 500 MHz, and 501 - 550 MHz are shown in solid-gray, solid-black and dashed-gray colors, respectively.
The values (normalized impedance) which lie between 0.5 and 2.0 in the Smith chart indicate that the impedance values of the antenna are closer to $Z_0$ in the 50 - 300 MHz range (except at 98 MHz); whereas they are farther almost over the rest of the band. To begin with, the high values of VSWR above 450 MHz were brought down close to 2.5, by increasing the high frequency cut-off from 555.6 MHz to 614.8 MHz. While the number of dipoles were increased from 17 to 22 which had flattened the VSWR spectrum within the intended bandwidth (i.e, OB) as mentioned earlier and also had reduced drastically the VSWR value at 98 MHz; the resultant VSWR is shown in Fig. \ref{fig:VSWRvardia} (gray colored dotted-line profile). The revised specifications of the LPDA are given in Table \ref{tab:LPDA_C8}. The above process is equivalent to making a LPDA whose design bandwidth is equal to $\approx$ 1.25 times as that of the OB, i.e. $\left(N=\frac{ln(1.25\,\beta)}{ln(1/\tau)}\right)$; the latter led $\tau$ and $\sigma$ to be equal to 0.89 and 0.06, respectively. 

Having done the above, the tests were continued further to bring down the VSWR below 2.0 throughout the OB. Results of various tests conducted, enabled us to understand that both transmission line and the half-wave dipoles that constitute the LPDA play an important role in deciding its overall impedance (\cite{stutzman1981antenna,ARRL1991}); the following subsections deal with them one by one.

\subsubsection{Selection of dipoles}
\label{subsubsec:sel_dipole}
\begin{figure}[!t]
\centering
\centerline{\includegraphics[width=0.5\textwidth]{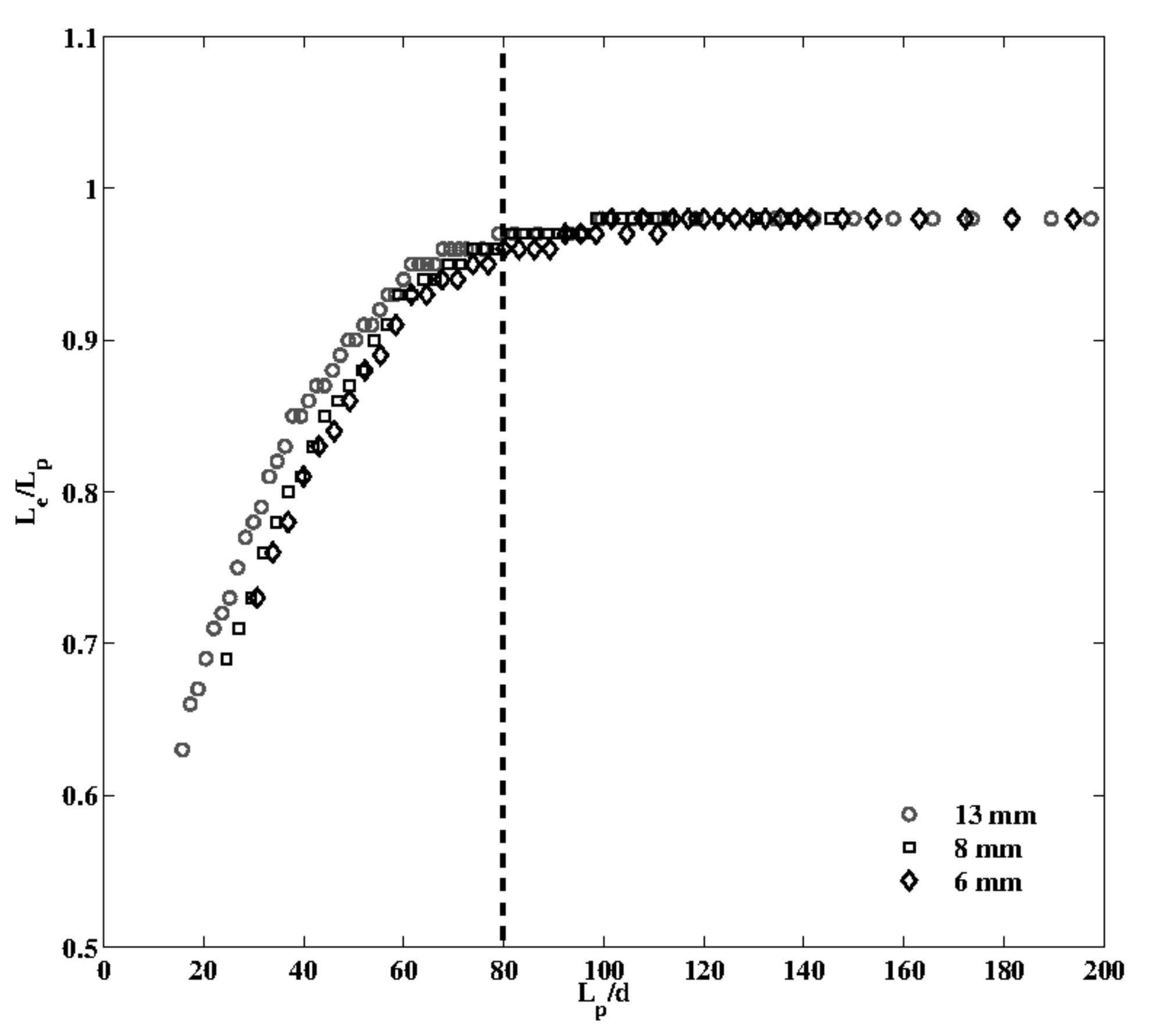}}
\caption{The ratio of electrical length ($ L_e$) to physical length ($L_p$) of a dipole is plotted against the ratio of its physical length to diameter ($d$). Measurements taken with Aluminum cylindrical tubes of outer diameter 13, 8, and 6 mm are shown with `circle', `square' and `diamond' symbols, respectively. A vertical dotted line is drawn at $L_p/d=80$, in order to show that $L_e/L_p$ approaches unity asymptotically beyond that.  \label{fig:LPDA_LebyLp}}
\end{figure}

\begin{figure}[!t]
\centering
\centerline{\includegraphics[width=0.8\textwidth]{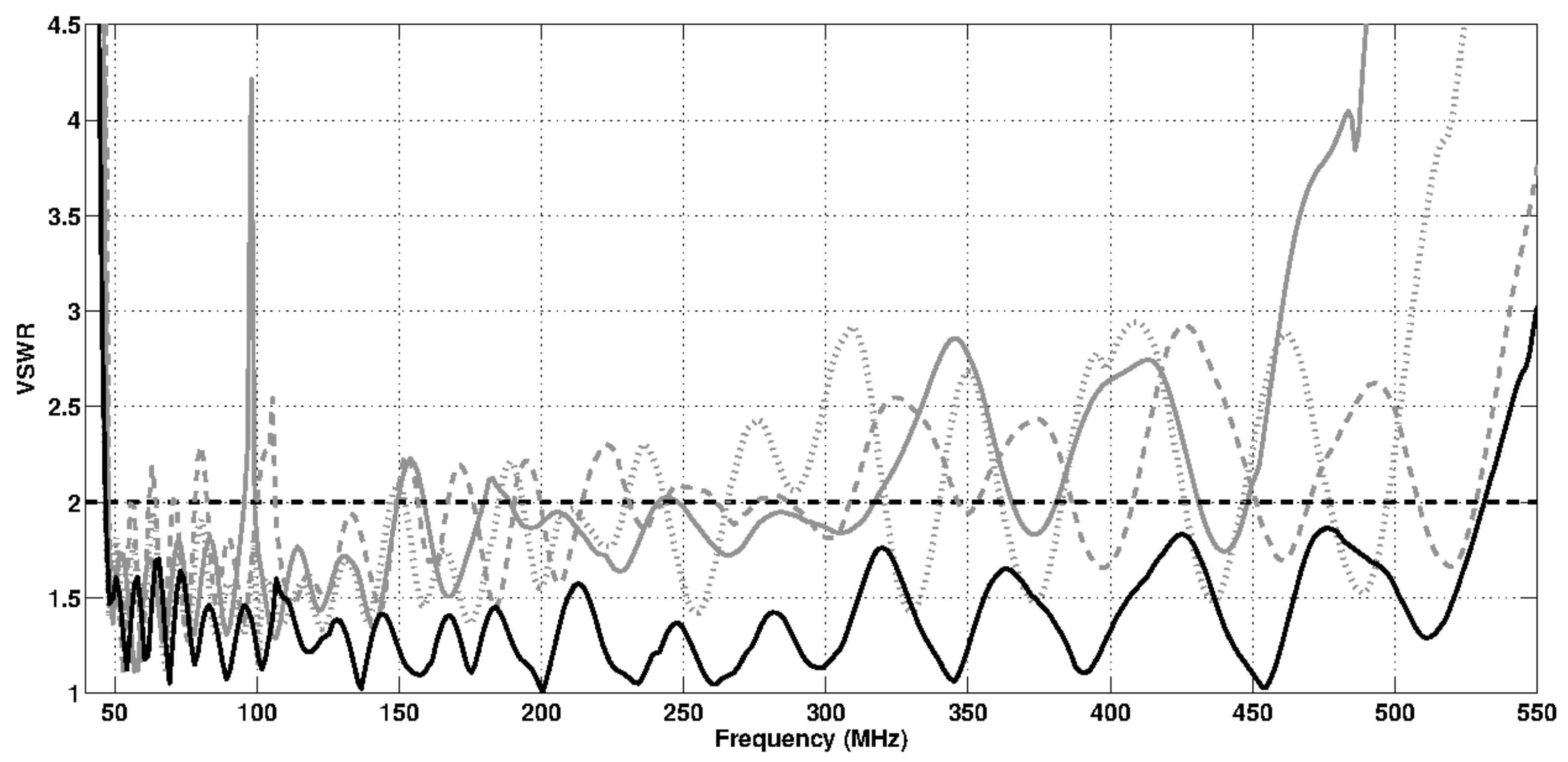}}
\caption{The VSWR profiles of the 50 - 500 MHz prototype LPDA. Solid-gray : Response with design specification given in Table \ref{tab:LPDA_C1}; Dotted-gray : Response with design specification given in Table \ref{tab:LPDA_C8}; Dashed-gray : Response after implementing the suggestion given in section \ref{subsubsec:sel_dipole}; Solid-black : Response after implementing the suggestions in sections \ref{subsubsec:sel_dipole} and \ref{subsubsec:trans_spacing}. \label{fig:VSWRvardia}}
\end{figure}

For broadband operation of a LPDA, choosing the right set of resonating dipole elements (i.e. with appropriate dimensions) is important since the location of active-centres of reception / transmission moves  with frequency  \cite{kraus1950}. The frequency at which a half-wave dipole element resonates depends on its electrical dimension which in turn is related to its physical dimension. In order to select the right set of dipoles that would work in the OB, we studied the half-wave dipole response alone for various physical lengths and diameters; Aluminum tubes of three different outer diameters, viz. 13, 8 and 6 mm were used to perform the tests; the result, $L_e/L_p$ Vs $L_p/d$, is shown in Fig. \ref{fig:LPDA_LebyLp}, where $L_e$, $L_p$, and d are the electrical length, physical length, and diameter of the half-wave dipole, respectively. 
In all the cases it can be seen that $L_e/L_p$ varies almost linearly with $L_p/d $ for values $\lesssim 80$; beyond that the electrical length of a half-wave dipole is almost equal to its physical length, because $L_e/L_p$ approaches unity asymptotically. The latter point was used effectively to fine-tune the impedance of the prototype LPDA : Sets of dipoles whose diameter satisfied the above criterion over the OB were selected to construct the LPDA. Aluminum tubes of different diameters (4 - 19 mm) available with us were used to cover the above bandwidth; Table \ref{tab:LPDA_C8} may be referred to know the exact diameter of dipoles used to construct the LPDA.
\begin{table}[htbp]
\caption{Revision to Table \ref{tab:LPDA_C1} plus diameter of dipoles}
\begin{center}
\begin{tabular}{ccccc}
\hline
\textbf{S. No.} & \textbf{Distance} & \textbf{Half-Dipole} & \textbf{Frequency} & \textbf{Dipole}  \\
\textbf{} & \textbf{from feed} & \textbf{Length ($\lambda/4$)} & \textbf{($f=c/\lambda$)} & \textbf{diameter} \\
\textbf & \textbf{(cm)} & \textbf{(cm)} & \textbf{(MHz)} & \textbf{(mm)} \\ \hline \
1. & 0.0 & 12.2 & 614.8 & 4 \\ 
2. & 3.0 & 13.7 & 547.5 & 4 \\ 
3. & 7.0 & 15.7 & 477.9 & 4 \\ 
4. & 11.0 & 17.7 & 423.9 & 4 \\ 
5. & 15.0 & 19.7 & 380.9 & 4 \\ 
6. & 20.0 & 22.2 & 338.1 & 6 \\ 
7. & 26.0 & 25.2 & 297.9 & 6 \\ 
8. & 32.0 & 28.2 & 266.2 & 6 \\
9. & 39.0 & 31.7 & 236.8 & 8 \\ 
10. & 47.0 & 35.7 & 210.3 & 8 \\ 
11. & 56.0 & 40.2 & 186.8 & 10 \\ 
12. & 67.0 & 45.6 & 164.3 & 10 \\ 
13. & 78.0 & 51.1 & 146.7 & 13 \\ 
14. & 91.0 & 57.6 & 130.1 & 13 \\ 
15. & 106.0 & 65.1 & 115.2 & 16 \\ 
16. & 122.0 & 73.1 & 102.6 & 16 \\ 
17. & 141.0 & 82.6 & 90.8 & 19 \\
18. & 162.0 & 93.1 & 80.6 & 19 \\ 
19. & 186.0 & 105.1 & 71.4 & 19 \\ 
20. & 212.0 & 118.0 & 63.5 & 19 \\ 
21. & 242.0 & 133.0 & 56.4 & 19 \\ 
22. & 276.0 & 150.0 & 50.0 &19 \\ 
\hline
\end{tabular}
\end{center}
\label{tab:LPDA_C8}
\end{table}
The VSWR profile of the LPDA, after the latest changes made, is shown in Fig. \ref{fig:VSWRvardia} (gray color dashed-line).

\subsubsection{Adjusting the spacing between transmission lines}
\label{subsubsec:trans_spacing}
\begin{figure}[!t]
\centering
\centerline{\includegraphics[width=0.5\textwidth]{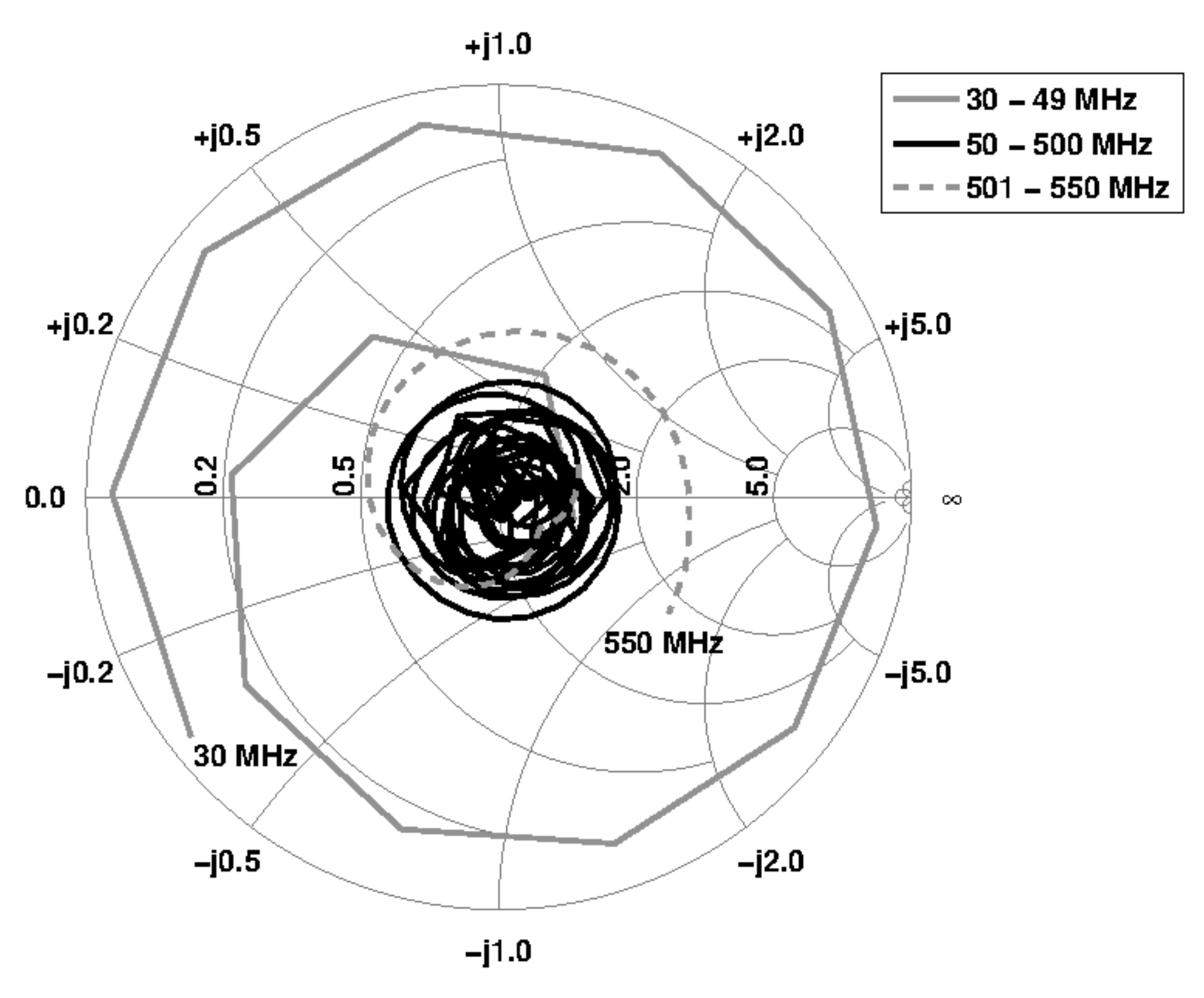}}
\caption{Smith chart of the prototype LPDA after introducing dipoles of different diameters and adjusting the inter-boom spacing; the corresponding parameter values are listed in Table \ref{tab:LPDA_C8}. \label{fig:LPDA_Smith_8}}
\end{figure}

The center-to-center spacing ($S$) between the transmission lines of the LPDA plays an important role \cite{ARRL1991} in deciding its impedance ($Z_{tr}$). Eqn. \ref{eq:spacing} describes the relationship between the input impedance ($Z_i$) of the LPDA, impedance of the nearest dipole ($Z_d$), $S$, and diameter of the transmission line ($D_{tr}$). 
\begin{equation}
\label{eq:spacing}
Z_i=Z_d\,\log\left({\frac{2S}{D_{tr}}}\right)
\end{equation}
The impedance of the dipole ($Z_d$) mentioned in Eqn. \ref{eq:spacing} depends upon the length ($l_n$) and diameter ($d_n$) of the dipole (Eqn. \ref{eq:dipoleZ}) \cite{ARRL1991}.
\begin{equation}
\label{eq:dipoleZ}
Z_d=120{\left(ln\left(\frac{l_n}{d_n}\right)-2.25\right)}
\end{equation}
Since we maintained the length-to-diameter-ratio of dipoles used in our design a constant ($\approx 80$), the actual values in $Z_d$ would have lain close to each other. Therefore, $Z_i$ in Eqn. \ref{eq:spacing} would directly depend upon $S$, because $D_{tr}$ was also kept constant in our design; this gave us an understanding that the increasing trend of $Z_i$ from low frequency cut-off to high frequency cut-off as seen from gray color dashed-line of Fig. \ref{fig:VSWRvardia}, could have been due to a progressive increase in $S$ (above the required value) toward the high frequency side; since the existing $S$ was a constant (20 mm throughout), the tests were continued by decreasing $S$ more towards the high-frequency side. As expected, this approach brought down the VSWR well below 2.0 throughout the OB, which resulted in overall matching of impedance (with $Z_0$) of the prototype LPDA. However, by varying the gradient of $S$, the configuration that yielded the lowest mean VSWR was identified. The one which had 6 mm close to the location of 500 MHz dipole and 60 mm close to the location of 50 MHz gave rise to 1.35 as mean VSWR. The solid-black profile shown in Fig. \ref{fig:VSWRvardia} was obtained after adjusting $S$; its corresponding Smith chart is shown in Fig. \ref{fig:LPDA_Smith_8};
one can see that the normalized impedance values lie well between 0.5 and 2.0 in the 50 - 500 MHz band. This shows that the LPDA can receive the radio waves in that frequency range effectively. Also, it must be emphasized that neither a 2:1 nor a 4:1 impedance transformer (BALUN) was used anywhere to match the impedance of the LPDA with $Z_0$, as is generally followed. Further, during the tests, it was found that similar VSWR response in the OB could be obtained by joining the transmission lines using an insulated-wire-stub of length ${\frac{\lambda_{max}}{8}}$; for compactness, the stub may be made into a coil; in such a case, the number of dipoles required to realize the OB with a coil-stub configuration would be lesser than the one without that. The reduction in number of dipoles depends upon the design parameters; it is equal to five for our prototype LPDA.

\section{Fabrication and characterization of the CLPDA}
\label{sec:fab_CLPDA}
\begin{figure}[!t]
\centering
\centerline{\includegraphics[width=0.5\textwidth]{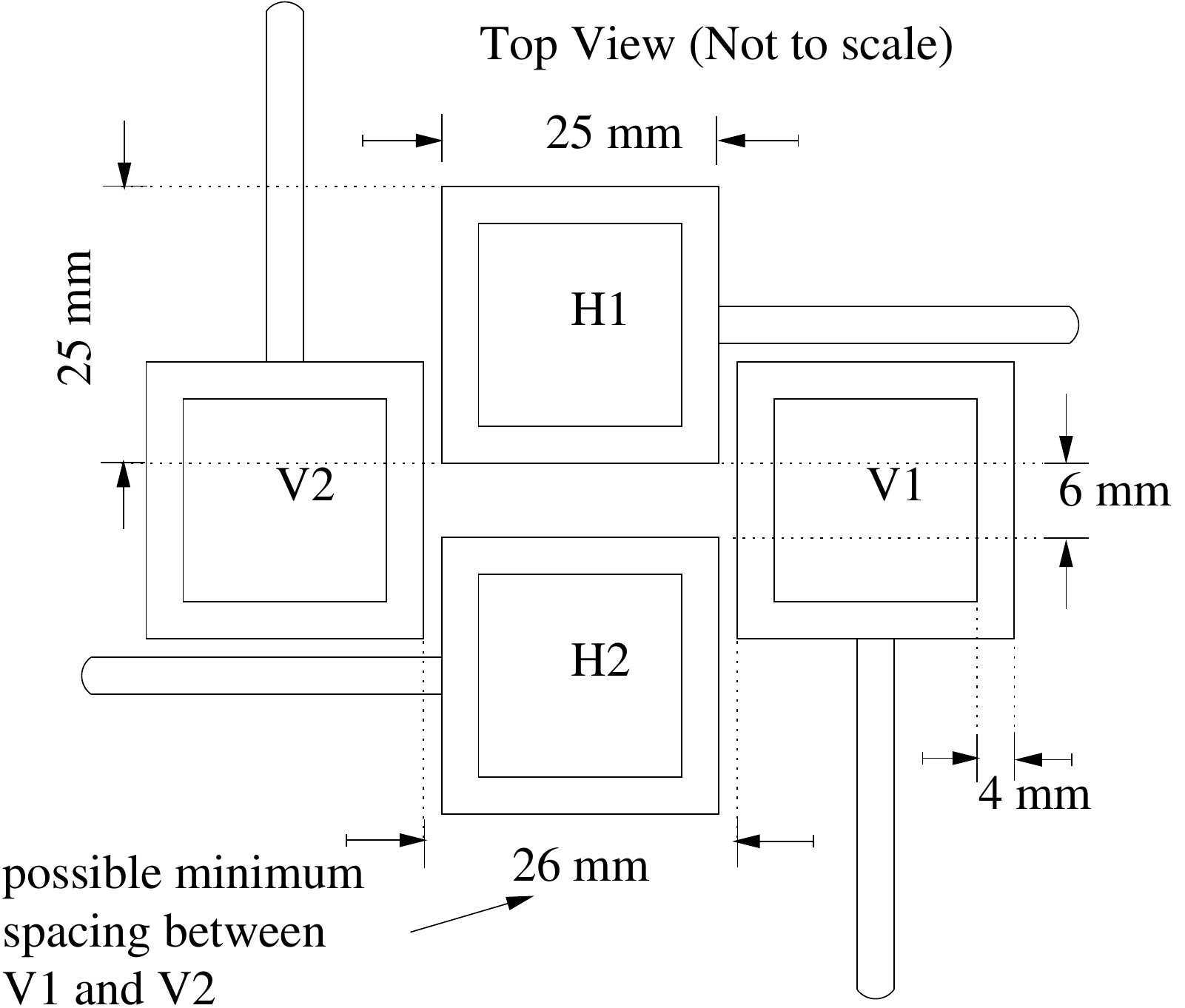}}
\caption{Schematic of the CLPDA with boom transmission lines; H1 and H2 / V1 and V2 pair holds the dipole elements oriented horizontally (X-pol.) / vertically (Y-pol.).  \label{fig:CLPDA_boom_schematic}}
\end{figure}

\begin{figure}[!t]
\centering
\centerline{\includegraphics[width=0.8\textwidth]{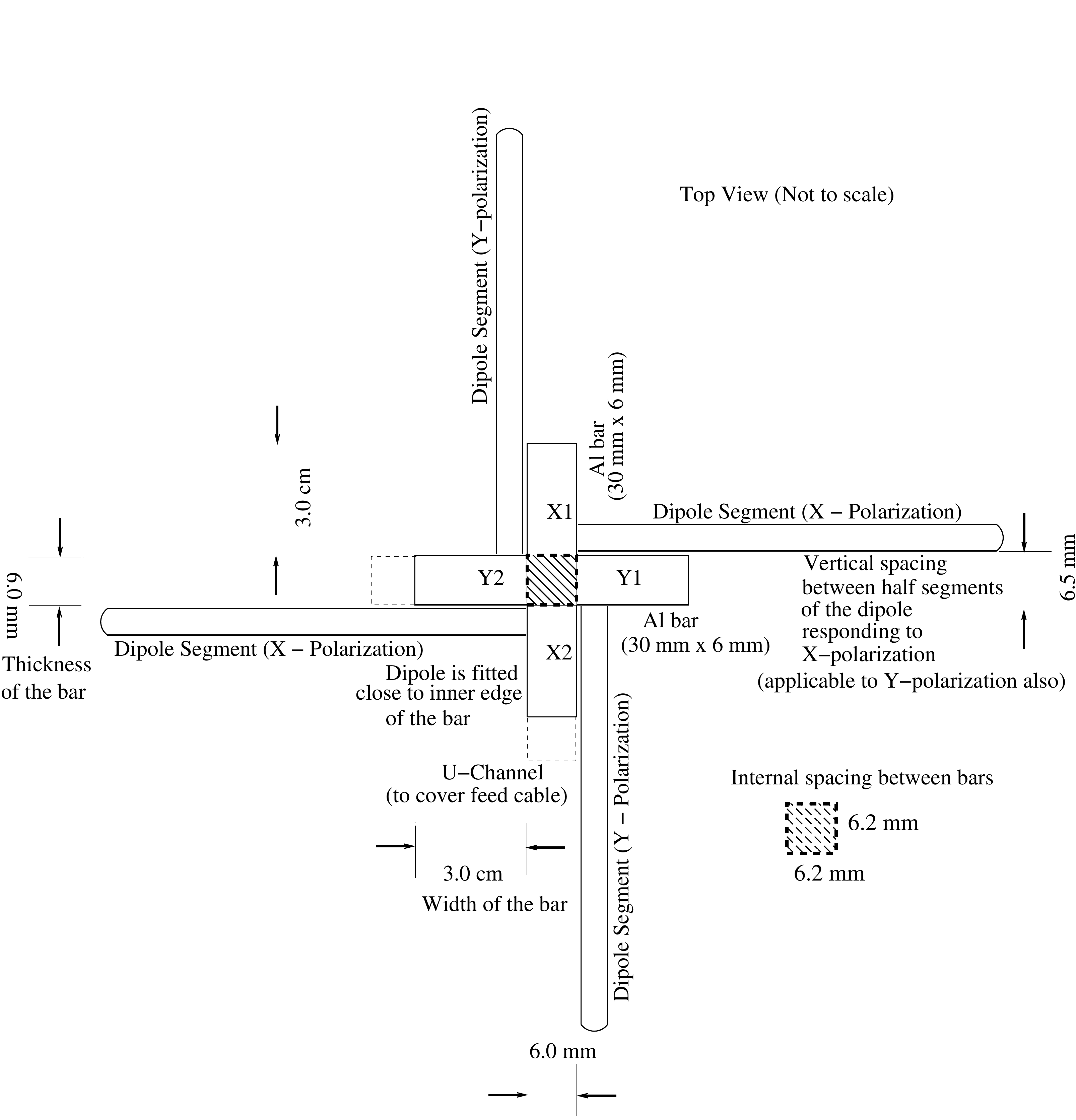}}
\caption{Schematic showing top view of the CLPDA with rectangular bar transmission lines; X1 and X2 / Y1 and Y2 pair holds the dipole elements oriented horizontally / vertically. \label{fig:CLPDA_schematic}}
\end{figure}

\begin{figure}[!t]
\centering
\centerline{\includegraphics[width=0.8\textwidth]{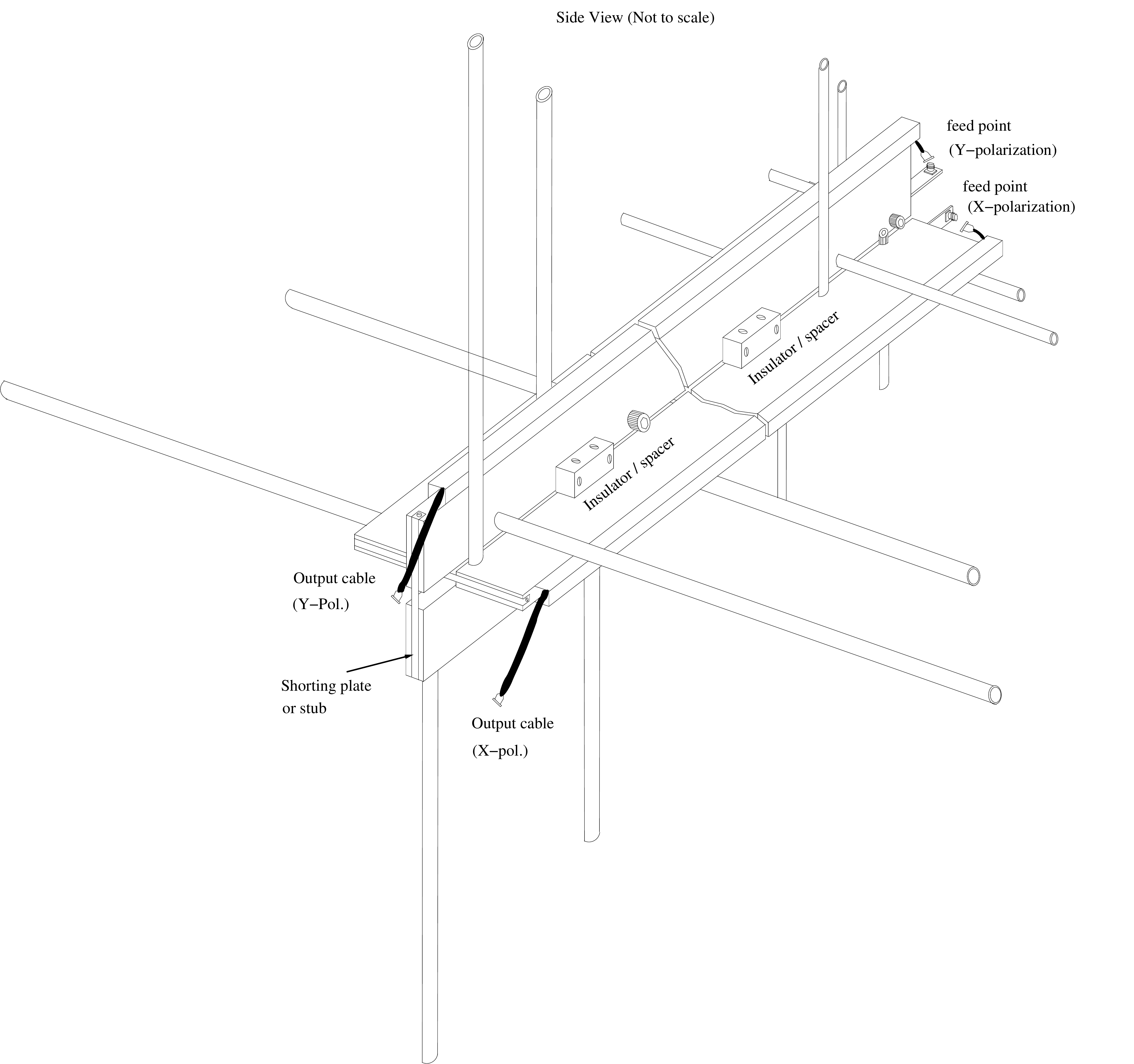}}
\caption{Same as Fig. \ref{fig:CLPDA_schematic}, but shown in side view. \label{fig:CLPDA_sideview}}
\end{figure}

\begin{figure}[!t]
\centering
\centerline{\includegraphics[width=0.8\textwidth]{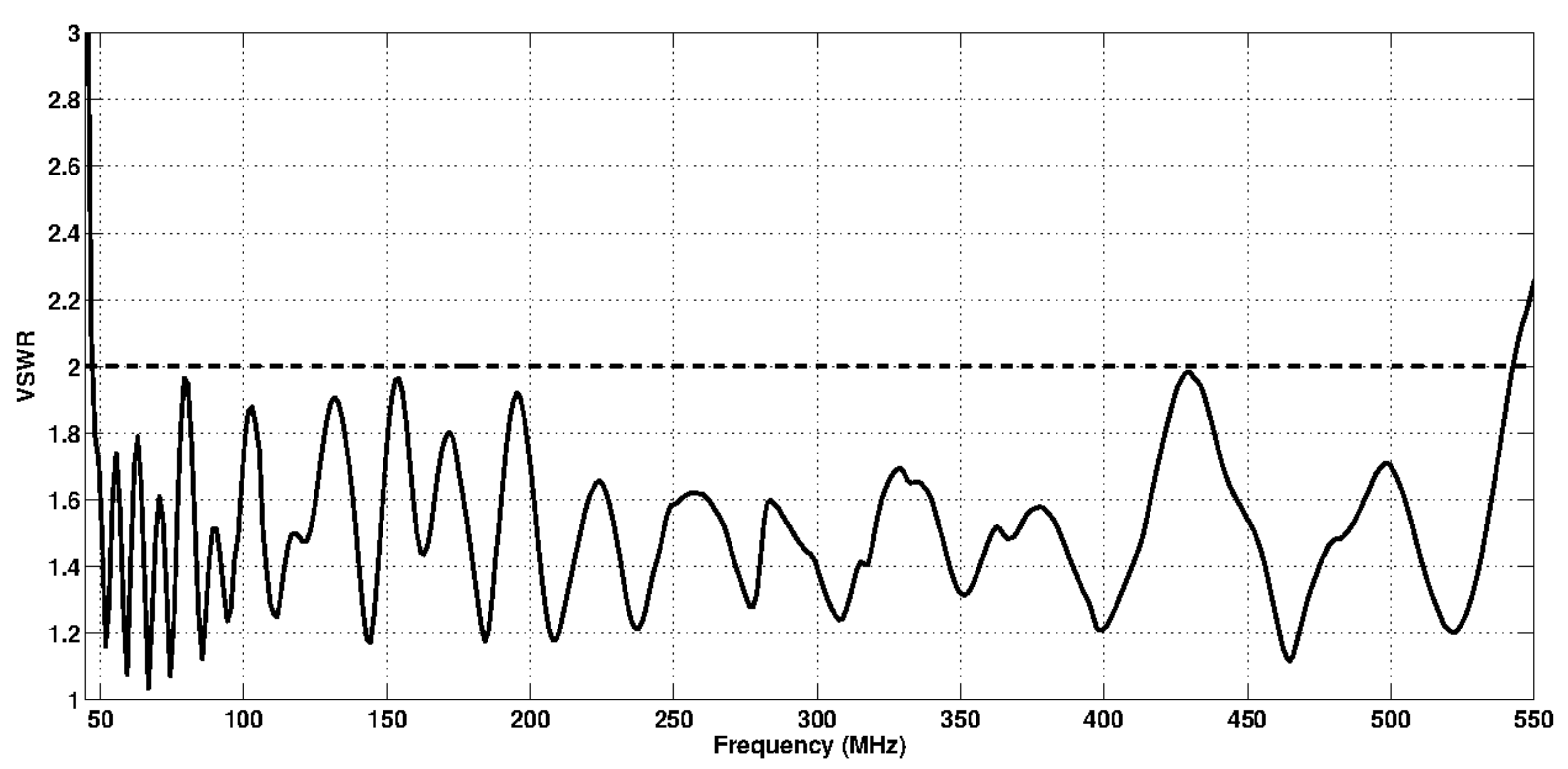}}
\caption{VSWR of the prototype CLPDA. \label{fig:CLPDA_VSWR}}
\end{figure}

\begin{figure}[!t]
\centering
\centerline{\includegraphics[width=0.6\textwidth]{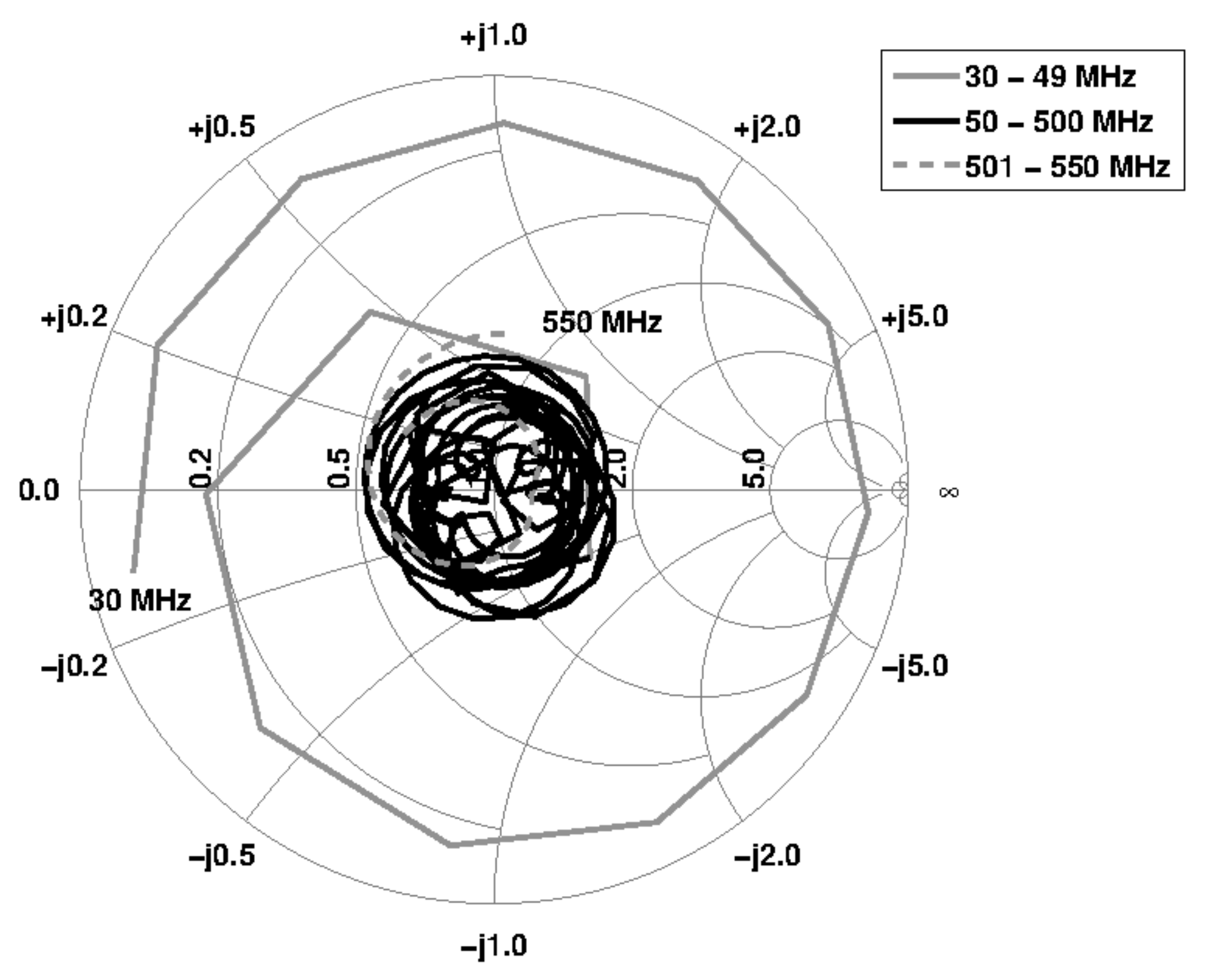}}
\caption{Smith chart of Fig \ref{fig:CLPDA_VSWR}. \label{fig:CLPDA_Smith}}
\end{figure}

\begin{figure}[!t]
\centering
\centerline{\includegraphics[width=0.4\textwidth]{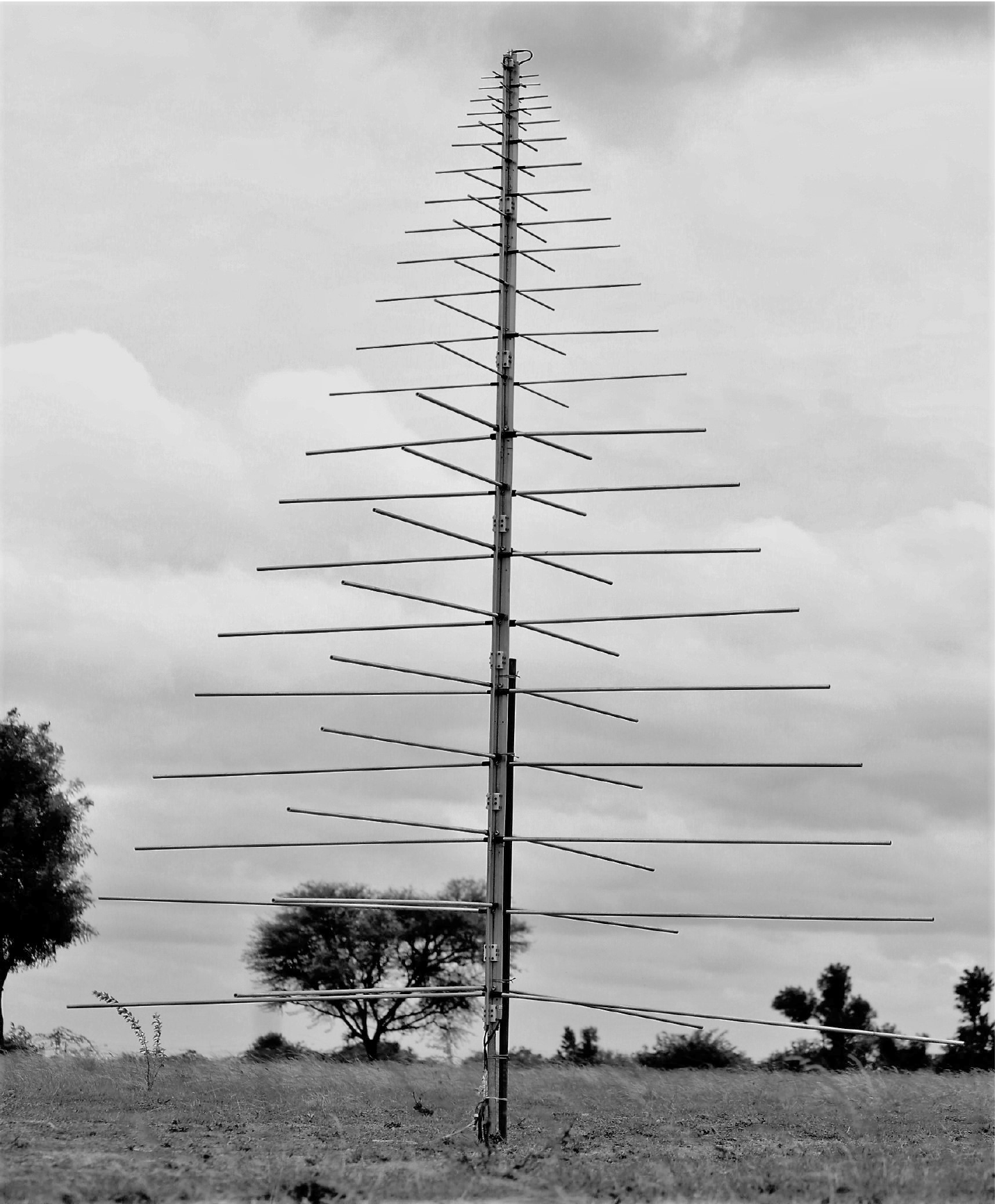}}
\caption{Photograph of the prototype CLPDA (foreground). \label{fig:CLPDA_photo}}
\end{figure}

As mentioned in the beginning of Section \ref{sec:DD_CLPDA}, a CLPDA is constructed by fitting two LPDAs with dipole-orientations orthogonal to each other with vertical axes being the same. Followed by the LPDA fabrication, a schematic was drawn (Fig. \ref{fig:CLPDA_boom_schematic}) as a part of the preparatory work to the CLPDA;
while drawing, it was realized that constructing a CLPDA using a boom type transmission line would not help because the transmission lines of the orthogonal components of the CLPDA should be separated by 6 mm at the top; an inspection of Fig. \ref{fig:CLPDA_boom_schematic} indicates that though the latter is achievable for one of the components (H1 and H2), it is not for the other (V1 and V2), since the dimension of the boom (25 mm $\times$ 25 mm) would be larger than the spacing required (i.e., 6 mm $\times$ 6 mm). Naturally, the dimension of the transmission line was constrained to be equal to 6 mm $\times$ 6 mm; yet, since the prototype LPDA was prepared using dipoles having different diameters (4 mm - 19 mm), the dimension of the transmission lines had to be increased from 6 mm $\times$ 6 mm to 6 mm $\times$ 30 mm (Fig. \ref{fig:CLPDA_schematic}) in order to fit those dipoles on to the CLPDA transmission lines.
Therefore, the transmission line pairs (X1-X2 and Y1-Y2) of the CLPDA were prepared out of slender (6 mm thickness $\times$ 30 mm width) Aluminum rectangular bars instead of the previously used 25 mm $\times$ 25 mm booms (with 4 mm wall-thickness). Rectangular bar was preferred to rectangular tube for better rigidity since it should run for a length of 3 m. Each transmission line pair was fitted with a `U' channel to the outer edge of one of the pair (X2 and Y2) to run the feed cable through it; two separate RG-58 coax cables, each having a length of 3.2 m, were used to tap the signal from the feed connectors fitted to X1-X2 and Y1-Y2 pairs, respectively (Fig. \ref{fig:CLPDA_sideview}); 
the latter pairs can be oriented such that the first one receives horizontal component and the second one receives the vertical component of the polarized signal falling on them. Before proceeding with the preparation of the CLPDA, we took into account the suggestions given in Pivnenko \cite{pivnenko2006log} which states that the spatial separation between the transmission lines of a LPDA / CLPDA at a particular location must be $\lesssim \frac{1}{100}$ times the wavelength of operation of the nearest dipole for restricting the electro-magnetic (EM) field vibrations within a very narrow cone angle; this was recommended to minimize the polarization leakage and thereby to detect the state of polarization, presumably with better precision. Therefore, an insulator of dimension 6.2 mm $\times$ 6.2 mm $\times$ 100 mm (width $\times $ height $\times $ length) was carefully fitted between two orthogonal components of transmission lines at several locations to ensure that the spacing between the transmission line at any location satisfies Pivnenko's criterion; also, the dipoles were fitted close to the inner edge of the rectangular bars so that the vertical spacing between the half segments that constitute the dipole is almost equal to the spacing between the transmission lines, which is equal to 6.5 mm (Fig. \ref{fig:CLPDA_schematic}), at the location of the dipole. Additional spacers were also fitted to vary the inter-transmission line spacing from about 6.5 mm to 60 mm. The location, inter-dipole spacing, diameter of the dipoles given in Table \ref{tab:LPDA_C8} were followed again in fabricating the prototype CLPDA; the VSWR was measured for it and the values were found to be below 2.0 throughout the OB. As was done earlier, the gradient in the inter-transmission line spacing was varied to identify the configuration that would give the lowest average-VSWR; the CLPDA configuration which had 6.5 mm spacing at the top and 30 mm at the bottom gave 1.52 as the lowest; the VSWR profile and the corresponding Smith chart are shown in Figs. \ref{fig:CLPDA_VSWR} and \ref{fig:CLPDA_Smith}, respectively.
 Its photograph is shown in Fig. \ref{fig:CLPDA_photo}.
A comparison of Fig. \ref{fig:CLPDA_VSWR} and solid-black profile in Fig. \ref{fig:VSWRvardia} shows that the VSWR of the CLPDA, and hence its impedance, differs from the prototype LPDA; this difference is most likely due to different transmission lines used to fabricate them as rest of the components used are the same for both antennas. Tests are being carried out to understand the effects of different types of transmission lines and their dimensions in obtaining the desired VSWR over the OB of interest. The results will be summarized in a different article and will be submitted later. Having developed, the prototype CLPDA was then taken for testing its reception characteristics.

\subsection{Radiation pattern measurements}
\label{sec:rad_patt}
\begin{figure}[!t]
\centering
\centerline{\includegraphics[width=1\textwidth]{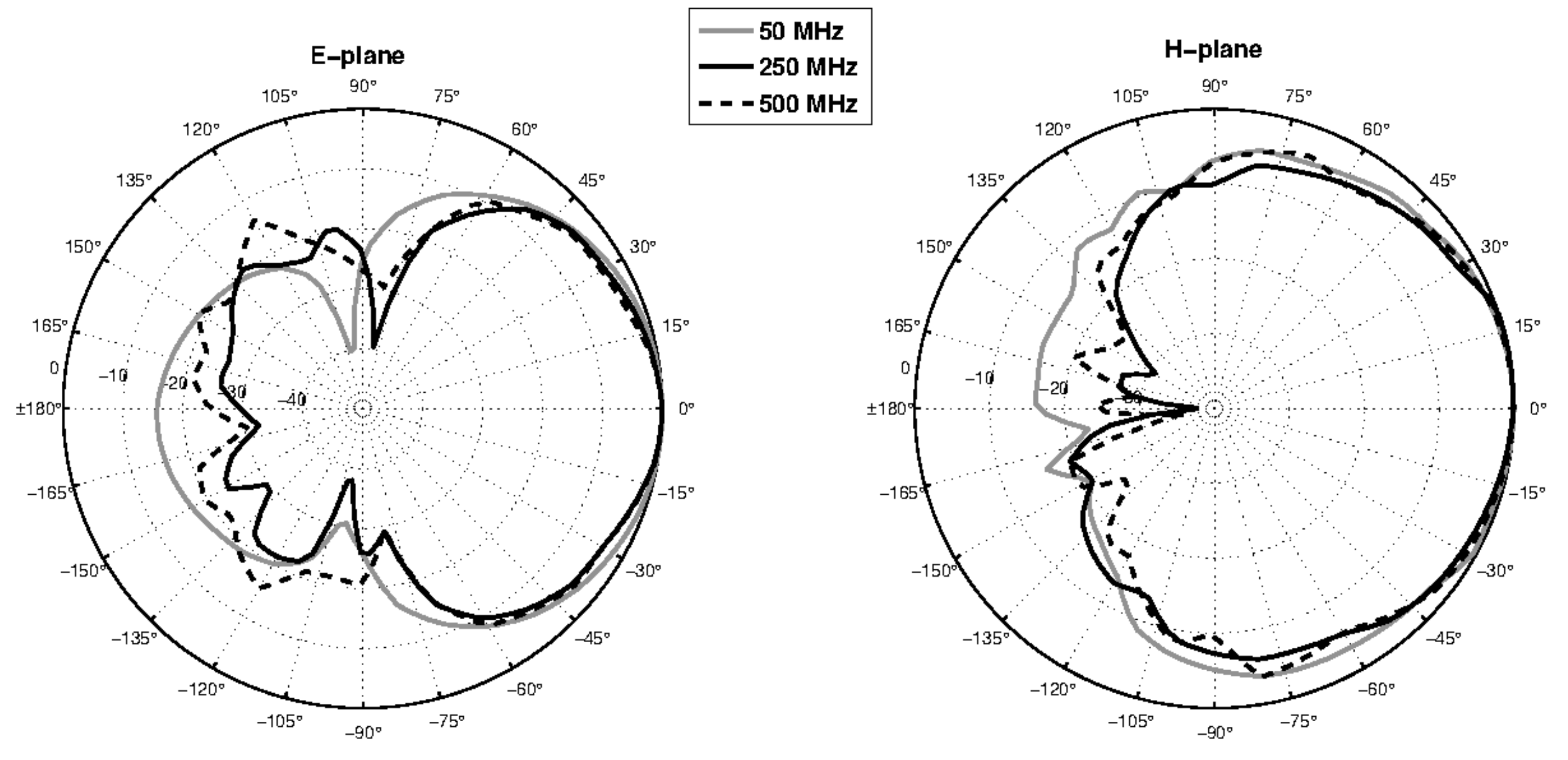}}
\caption{Measured E-plane (left panel) and H-plane (right panel) radiation patterns at 50, 250 and 500 MHz.}
\label{fig:CLPDA_EH}
\end{figure}

The far field radiation pattern was measured by keeping the transmitting and receiving antennas 40 m apart because the theoretical far field distance  ($R>\frac{2D^2}{\lambda}$) limit is about 30 m. The designed antenna was used as a receiver and a LPDA with known transmission characteristic was used as the transmitter. For the measurement of E-plane radiation pattern, both transmitting and receiving antennas were oriented horizontally and the receiver antenna was rotated in the azimuthal plane from $0^\circ$ reference position to $360^\circ$; the readings were noted down for every $15^\circ$. Same procedure was repeated for measuring the H-plane pattern but with both antennas oriented in vertical position. The measured average half power beam width (HPBW) of E-plane and H-plane are $65^\circ$ and $100^\circ$, respectively. Other antenna parameters such as gain, effective aperture area, etc. were deduced from the above and are listed in Table \ref{tab:LPDA_param}. These parameters were found to be constant almost throughout the OB. Fig. \ref{fig:CLPDA_EH} shows the radiation pattern measured at 50 (`solid-gray'), 300 (`solid-black') and 500 MHz (`dashed-black') for E-plane and H-plane, respectively.
\begin{table}[htbp]
\caption{Measured and deduced CLPDA parameters}
\begin{center}
\begin{tabular}{c l l}
\hline
\textbf{S. No.} & \textbf{Parameter} & \textbf{Value} \\ \hline
\\
1. & HPBW (E) & $\approx65^\circ$ \\  
2. & HPBW (H) & $\approx100^\circ$ \\  
3. & Gain & $\approx$ 6.6 dBi \\  
4. & Effective Collecting Area & $\approx 0.4\,\lambda^{2}$ \\ 
5. & Front to back ratio & $\approx$ 30 dB \\
6. & Side lobe ratio & $\approx$ -24 dB \\ 
7. & Polarization Cross-Talk & $\approx $ 30 dB \\
\\
\hline
\end{tabular}
\end{center}
\label{tab:LPDA_param}
\end{table}

\subsection{Estimation of Polarization Cross-talk}
\label{subsec:pol_crosstalk}
\begin{figure}[!t]
\centering
\centerline{\includegraphics[width=1\textwidth]{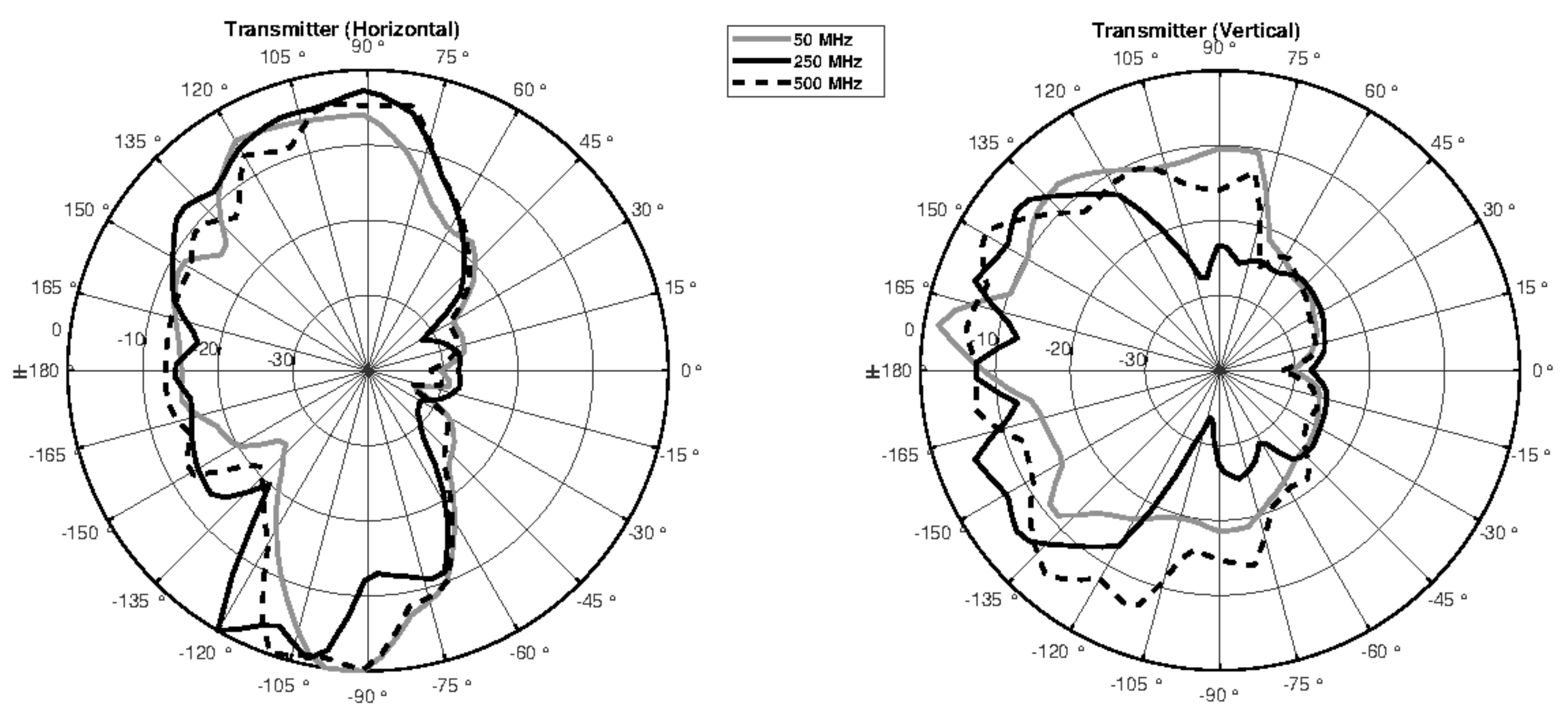}}
\caption{Polarization cross-talk or isolation pattern of the CLPDA at 50 (solid-gray), 250 (solid-black), and 500 MHz (dotted-black). Left : Transmitter kept in horizontal orientation; Right : Transmitter kept in vertical orientation.}
\label{fig:CLPDA_isolation_pattern}
\end{figure}

\begin{figure}[!t]
\centering
\centerline{\includegraphics[width=0.7\textwidth]{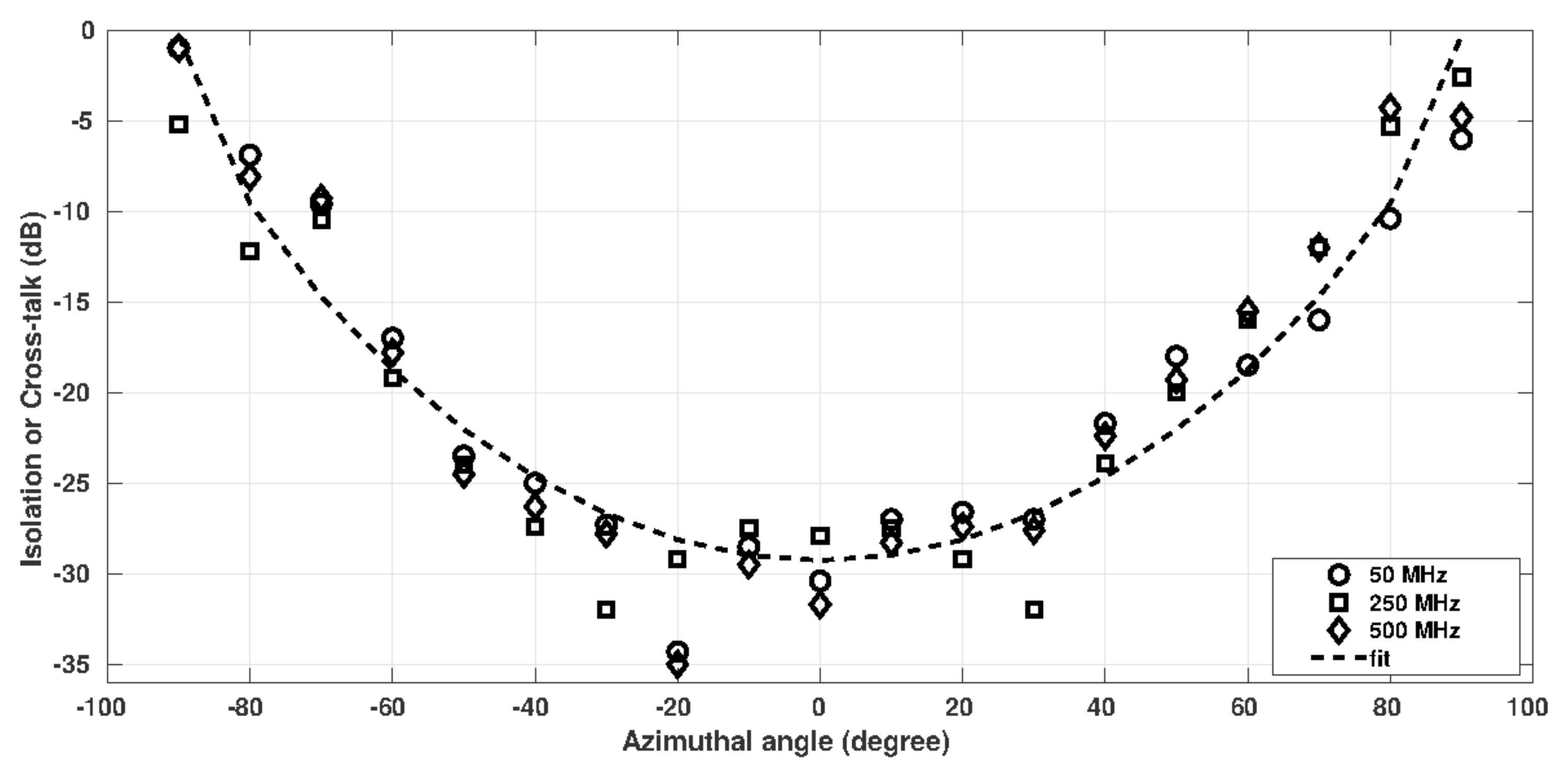}}
\caption{Measured polarization cross-talk of the CLPDA as a function of azimuthal angle when the transmitter was kept in horizontal orientation. The values at 50, 250, and 500 MHz are shown with open circle, square, and diamond symbols, respectively. A custom fit ($-a\,{cos^b}\theta - c$) to the average cross-talk values is shown with dash marks. The best fit ($\chi^2 \approx$ 0.93 and rms error $\approx$ 2.5) was obtained when the coefficients $a$, $c$ and power index $b$ were equal to $\approx$ 30, 0.3, and 0.65, respectively. The latter gives $\approx$ -30, -29, -27, -24, -19, -13 and 0 dB as the average cross-talk values at position angles $0^\circ, 15^\circ, 30^\circ, 45^\circ, 60^\circ, 75^\circ$ and $90^\circ$, respectively.}
\label{fig:CLPDA_avg_H_isolation}
\end{figure}

\begin{figure}[!t]
\centering
\centerline{\includegraphics[width=0.8\textwidth]{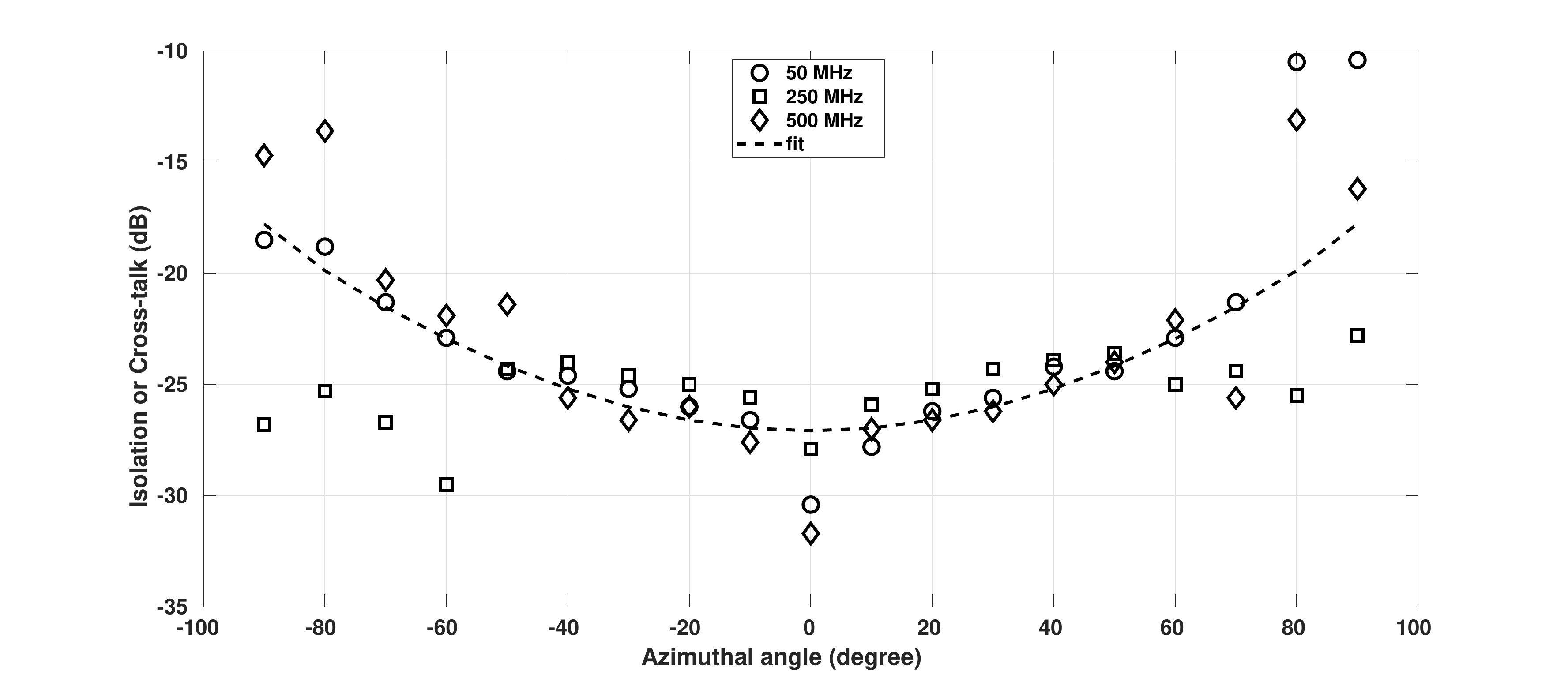}}
\caption{Same as Fig. \ref{fig:CLPDA_avg_H_isolation} but when the transmitter was kept in vertical orientation. The best fit ($\chi^2 \approx$ 0.82 and rms error $\approx$ 2.0) was obtained when the coefficients $a$, $c$ and power index $b$ were equal to $\approx$ 9.3, 17.8, and 0.85, respectively. The average cross-talk values are $\approx$ -27, -27, -26, -25, -23, -21 and -18 dB at $0^\circ, 15^\circ, 30^\circ, 45^\circ, 60^\circ, 75^\circ$ and $90^\circ$, respectively.}
\label{fig:CLPDA_avg_V_isolation}
\end{figure}

Polarization pattern of an antenna depends upon its geometry \cite{stutzman1981antenna}. 
For instance, a helix is circularly polarized whereas a dipole is linearly polarized. Since the primary objective of this study is to measure the circularly polarized radio emission from the Sun using a CLPDA frontend system, restricting the E- \& H-fields, within a narrow region about their respective mean positions of vibration is important. Otherwise, one of the fields will spill over into the other. The magnitude of spill-over determines the uncertainty involved in any parameter deduced from the polarization measurements. The parameter that quantitatively describes this spill over is called as the polarization cross-talk or isolation
(\cite{wakabayashi1999circularly,pivnenko2006log}), a measure of received power corresponding to one sense of polarization by an antenna when is exposed to a polarized radiation of the other or of opposite sense. As mentioned above, minimization of this parameter is expected to improve the accuracy with which the degree of circular polarization (DCP) is determined; and consequently the determination of magnetic field strength from DCP. In order to determine the cross-talk, the transmitter was kept in both horizontal and vertical orientations successively, and the signal strengths were measured with both horizontal and vertical arms of the CLPDA. Fig. \ref{fig:CLPDA_isolation_pattern} shows the cross-talk or isolation pattern of the CLPDA at different frequencies; the left / right panel corresponds to the CLPDA response when the transmitter was kept in horizontal / vertical orientation; the profiles show a minimum of -36 dB (at 500 MHz) and a maximum of -20 dB (at 50 MHz) within the $\pm 60^\circ$ azimuthal angle from the reference position ($0^\circ$, i.e., the direction of maximum radiation of the transmitter). 
Since it was planned to observe the sources for about 3 hours on either side of the local meridian and for different declinations, the average cross-talk as a function of azimuthal angle was determined for both orientations (Horizontal and Vertical) of the transmitter; custom fit to the cross-talk values are shown in Figs. \ref{fig:CLPDA_avg_H_isolation} and \ref{fig:CLPDA_avg_V_isolation}, respectively.

\begin{figure}[!t]
\centering
\centerline{\includegraphics[width=0.5\textwidth]{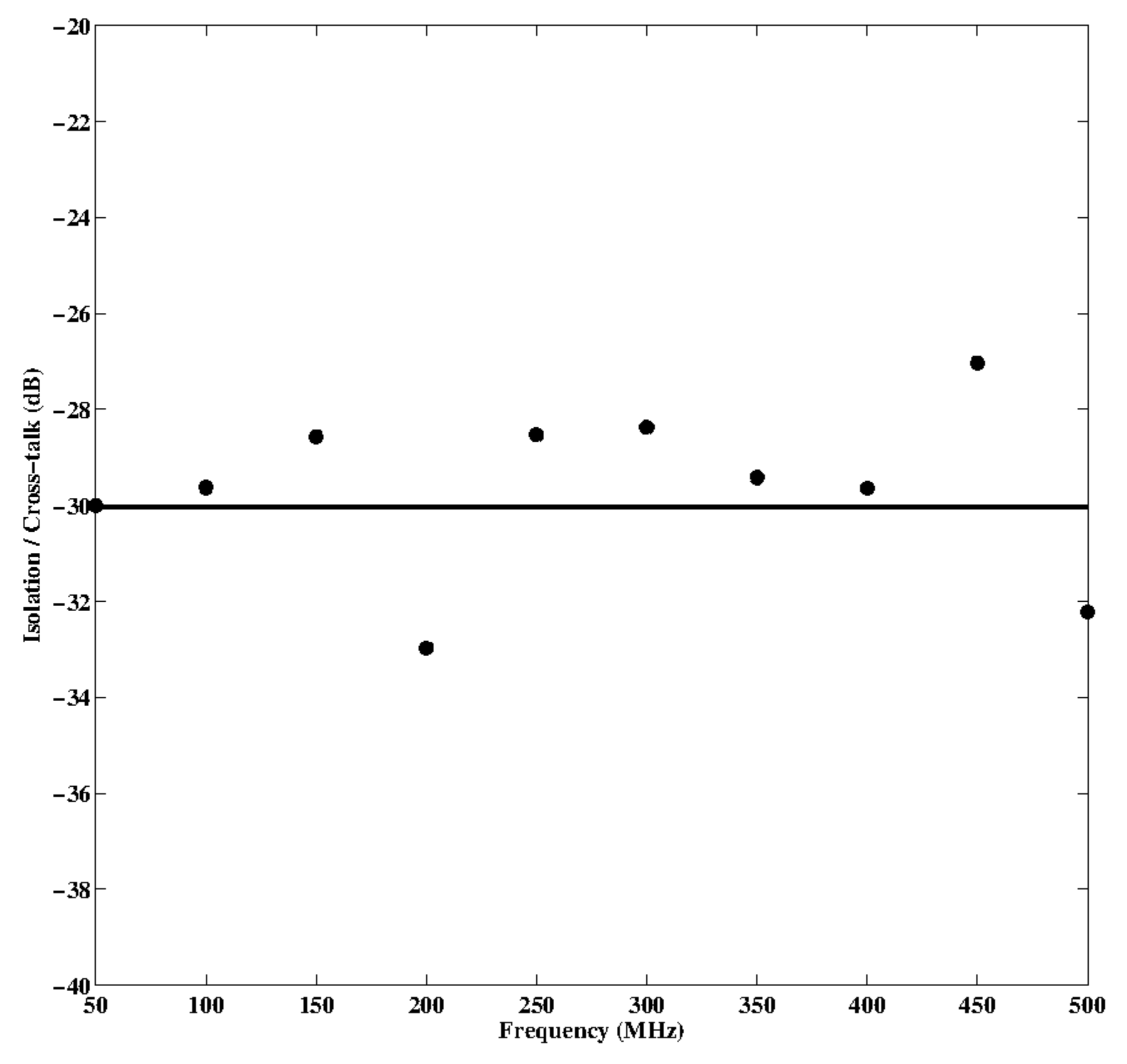}}
\caption{Isolation / Cross-talk of the CLPDA at $0^\circ$ azimuthal angle Vs frequency. The `filled circle' symbols correspond to the measured cross-talk level at different frequencies and the 'solid-black' line represents the mean value.}
\label{fig:CLPDA_crosstalk}
\end{figure}

The cross-talk values at $0^\circ$ position alone were plotted as a function of frequency and is shown in Fig. \ref{fig:CLPDA_crosstalk}; its average value is about -30 dB; this is about 10 dB lower than that are available commercially; majority of the latter use booms as transmission lines.

\section{Setting up of the Spectro-polarimeter}
\label{sec:SP_setup}
\begin{figure}[!t]
\centering
\centerline{\includegraphics[width=1\textwidth]{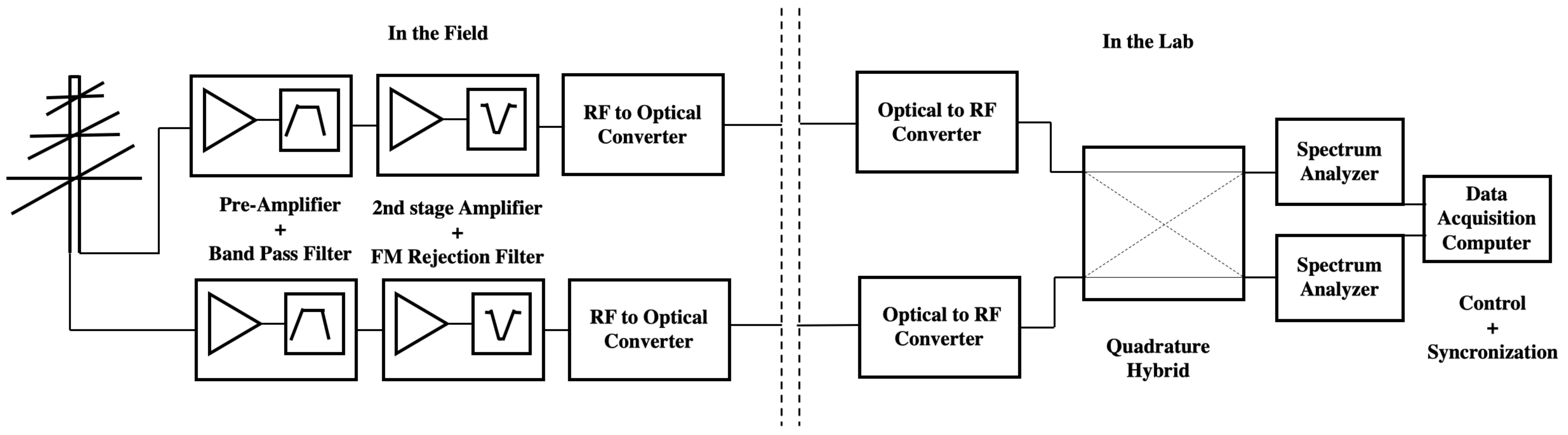}}
\caption{Block diagram of the spectro-polarimeter setup}
\label{fig:SP_blockdiagram}
\end{figure}

The new CLPDA was connected to an Analog frontend receiver and a digital backend receiver (Spectrum Analyzer) for setting up the Spectro-Polarimeter (SP). Its block diagram is shown in Fig. \ref{fig:SP_blockdiagram}. 

\subsection{Analog frontend \& digital backend receivers}
\label{subsec:ana_dig_rcvr}
The signals received by the two orthogonal arms of the CLPDA were fed into two analog front-end receivers. The first component of the analog receiver was a low-noise amplifier (MAN-1LN of M/s. Mini Circuits; noise figure 
$\approx $ 3 dB) which could amplify the signal by $ \approx $ 30 dB. Then the signal was passed through a high-pass filter whose cut-off frequency is $\approx $ 50 MHz; this was kept to eliminate the high level of radio frequency interference (RFI) that often prevail below 50 MHz at the Gauribidanur observatory. Also a band-stop filter (Insertion loss = 0.3 dB) was used to reject FM band (88 - 108 MHz). The filtered signal was amplified again using a second stage amplifier (similar to the low-noise amplifier) for ensuring the power level at the input of RF-to-optical converter (RF-OPC) to be optimum. The RF-OPC converts the RF signal using a low-power (+4 dBm at 1310 $\pm$ nm) class-1 uncooled Multi-Quantum Well (MQW) type LASER; it has a signal flatness of $\pm$ 1 dB and a dynamic range of about 100 dB (from -85 dBm to +15 dBm). It can safely take a maximum of 10 dBm RF signal in the 2 MHz - 3 GHz frequency range at its input port (input impedance $ \approx 50 ~\Omega $). The RF signal level was maintained to be $\lesssim$ -50 dBm over the OB to provide sufficient allowance for the increase in signal strength that would crop up at the time of high solar activity. The Carrier to Noise Ratio (CNR), Gain Stability, Spurious Free Dynamic Range (SFDR), and Noise Figure (NF) of the RF-OPC are $\approx$ -60 dB, 0.25 (over 24 hours), -70 $\rm dB/Hz^{2/3}$ and 18 dB, respectively. The converted signals were brought to the laboratory via optic fiber cables. 

In the lab, the original RF signals were retrieved from optical signals using an Optical-to-RF Converter (OP-RFC); the latter can convert the Optical signal into RF using a photo-diode having a spectral response of 1100 nm - 1650 nm. The signal flatness of this system is $\pm$ 1 dB over 2 MHz - 3 GHz range. The overall loss in the signal strength during the conversion from RF to optical and back to RF, including that of the OFC splicing joints ($\approx$ 0.5 dB per splicing) and the connectors ($\approx$ 1 dB per connector), is $\approx$ 20 dB. But since the intrinsic gain of the combined system is about 20 dB, the above conversion loss was compensated. The RF signals retrieved were then given to the inputs of a Quadrature Hybrid (QH; Insertion loss = 1.5 dB), a product of M/s. Synergy Microwave Co. 

QH is a four port RF device having two input and two output ports. Possessing two cross-over transmission lines as its internal circuitry, this device is used to couple two RF signals fed at its input ports to produce two different signals at its output ports. The signal fed at each input port is split initially into two components whose amplitudes are nearly equal but with a $90^\circ$ phase difference; the unaltered signal is called in-phase component whereas $90^\circ$ phase-shifted signal is called the quadrature component. The in-phase signal from the first input port and the quadrature component from the second input port are added at the first output port whereas the other two components contribute to the output of the second output port.

\begin{figure}[!t]
\centering
\centerline{\includegraphics[width=0.7\textwidth]{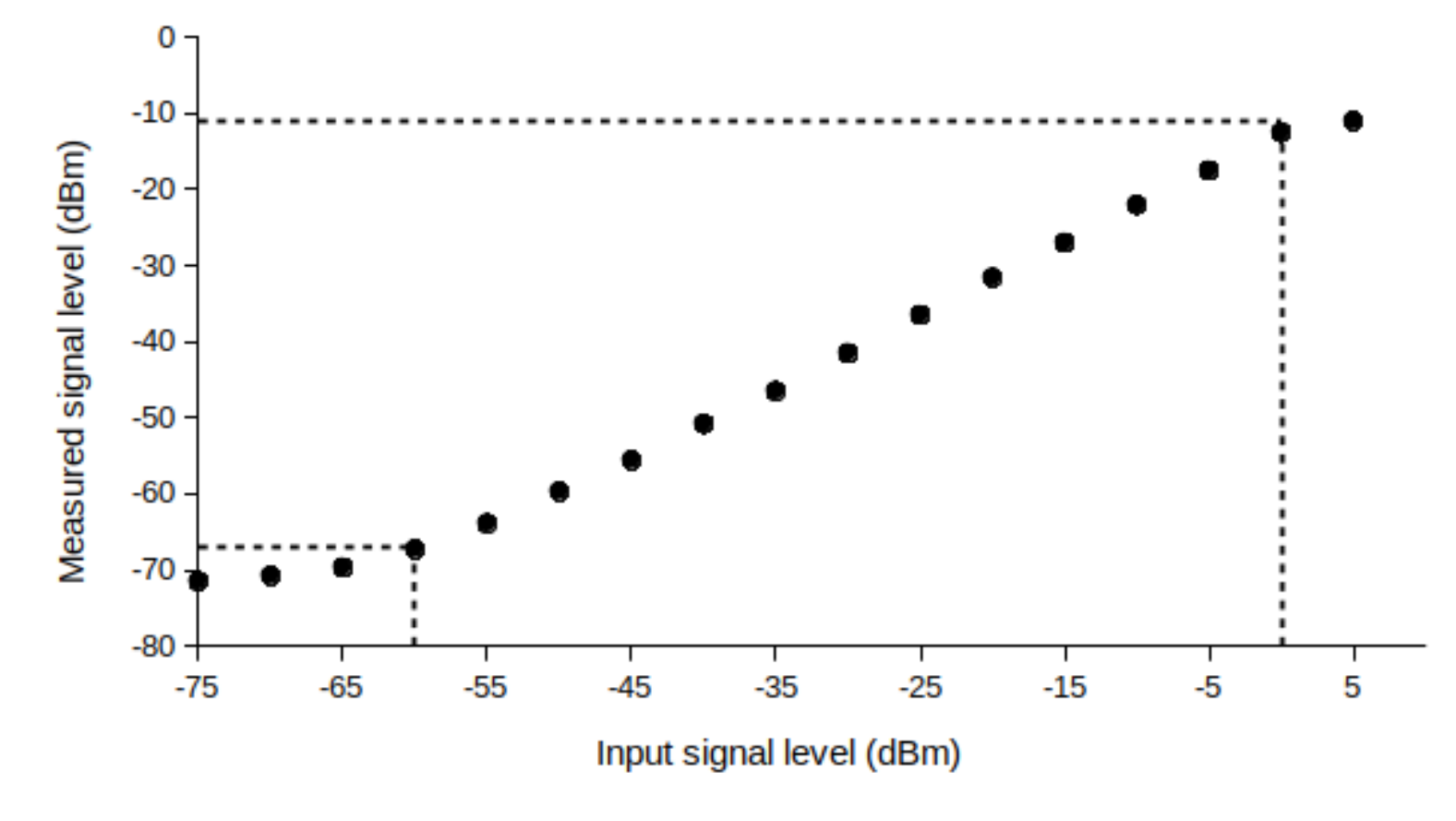}}
\caption{Dynamic range of the SA. It can be operated safely between -65 dBm and 0 dBm because of its linear response in that range. The offset between the input and measured signal amplitudes is due to the intrinsic attenuation ($\approx$ 10 dB) of the SA.}
\label{fig:dr}
\end{figure}

The outputs of the QH were fed into two conventional spectrum analyzers (SAs). The E4411B SAs (product of M/s. Agilent Technologies) were used to generate the spectrum of the input signals. The instrument works in the frequency range 9 kHz - 1.5 GHz with a frequency accuracy and resolution bandwidth range of $\pm$ 2 kHz, and 1 kHz - 5 MHz, respectively. The maximum dynamic range is about 80 dB. Tests were performed in the lab by varying the amplitude of a band-limited (50 - 500 MHz) noise to determine its dynamic range; the result is shown in Fig. \ref{fig:dr}; it can be seen that the response is linear in the -65 dBm to 0 dBm range; therefore, its effective dynamic range is $\approx$ 65 dB. 
The SA takes about 5 ms to 2000 s to sweep and cover the desired bandwidth based on the other parameter settings (that are programmable). A program was written in VEE pro \cite{AgilentVEE} (version 6.0; a GUI-based programming language that was supplied by the manufacturer and optimized for building test and measurement applications, and programs with operator interfaces) to operate, control and acquire the data from the SA. The program makes the SA to sweep and cover the OB and transfer the spectral information to the PC using GPIB (General Purpose Information Bus) interface. The GPIB works under IEEE-488 bus architecture. It has eight data lines (to transfer information), three handshake lines (to transfer information across data lines), five bus management lines (for general control and coordination of bus activities), and eight ground lines (for shielding and signal returns). 

\begin{figure}[!t]
\centering
\centerline{\includegraphics[width=0.8\textwidth]{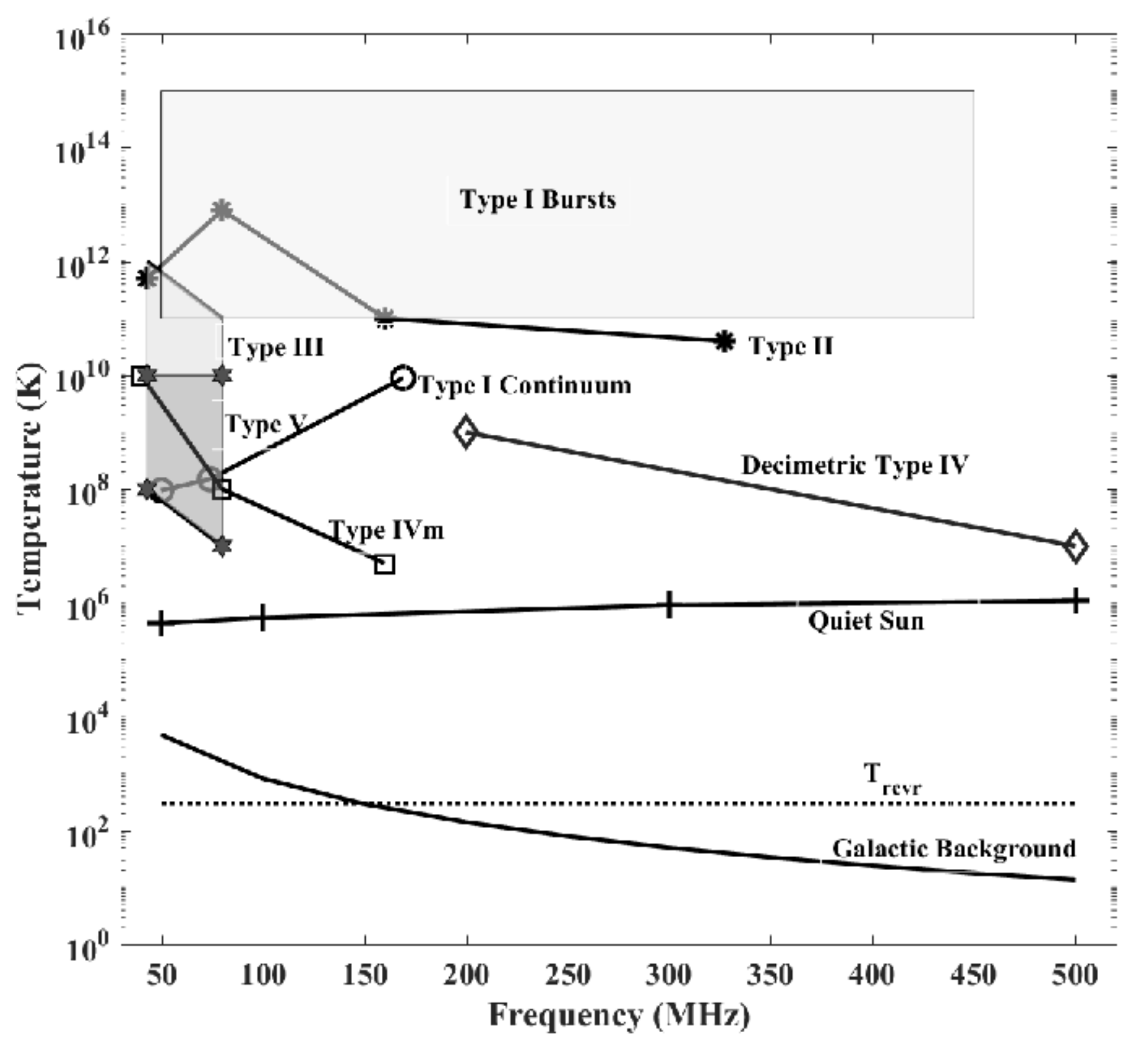}}
\caption{Brightness temperature of different types of solar radio bursts in comparison with $T_{rcvr}$ as a function of frequency.}
\label{fig:sun_bg_burst_tb}
\end{figure}

Both SAs were initialized using the GPIBs which set the observational parameters such as start frequency (50 MHz), end frequency (500 MHz), sweep time (4 ms), no. of data points (401),  BW resolution ($\approx$ 1 MHz), number of spectra to be averaged onboard (10 records), etc. The data transfer from the SA to the PC takes about 240 ms. An initial software trigger was given to start the frequency-sweep to synchronize the SAs. The flux density detectable by the receiver ($S_{rcvr}$) alone was calculated using Eqn. \ref{eq:RFD}.
\begin{equation}
\label{eq:RFD}
S_{rcvr} = \frac{2 k}{A_e} \frac{K_s~ T_{rcvr}}{\sqrt{\Delta \nu~ \Delta t~ n}}
\end{equation}
In Eqn. \ref{eq:RFD}, $k$ and $A_e$ are Boltzmann Constant and effective aperture ($0.4~\lambda^2$), respectively; while $K_s$, $T_{rcvr}$, $\Delta \nu$, $\Delta t$ and $n$ are sensitivity constant, receiver noise temperature, integration bandwidth, integration time, and number of records averaged onboard, respectively. The corresponding values for the receiver system are $\approx$ 1.0, 290 K, 1.1 MHz, $10~\mu s$, and 10, respectively. The latter values, at 50 and 500 MHz, give a flux density of $5.3 \times 10^3$ Jy (or 0.53 sfu, where 1 sfu = Solar Flux Unit = $10^4$ Jy) and $5.3 \times 10^5$ Jy (53 sfu), respectively, for the SP receiver. The values of brightness temperature of different types of solar radio bursts reported in the literature (\cite{Kundu1980,mclean1985solar,Aschwanden2006}) are plotted along with $T_{rcvr}$ to show that the present SP would enable us to detect almost all kinds of bursts (Fig. \ref{fig:sun_bg_burst_tb}).

\subsection{Detecting circularly polarized radio waves using a QH}
\label{subsec:CP_QH}
Let us assume that $\sqrt{2}A_o$ is the amplitude of a circularly polarized radio wave that is falling on to a CLPDA. The x- and y- components of the wave received by the two orthogonal components can be written as, 
\begin{equation}
\label{eq:CPW_xcomp}
A_{x}(t) = A_{ox}\exp^{i\delta_x}\exp^{i\omega t}
\end{equation}
\begin{equation}
\label{eq:CPW_ycomp}
A_{y}(t) = A_{oy}\exp^{i\delta_y}\exp^{i\omega t}
\end{equation}
Here, $A_{ox}$, $A_{oy}$, $\delta_x$, and $\delta_y$ are the instantaneous amplitude and spatial frequency of the wave along the x and y axes, respectively. The term $ \omega$ represents the angular frequency of the wave. As mentioned earlier, the two outputs of the QH, viz. $A_{QH}^{0^\circ}(t)$ and $A_{QH}^{90^\circ}(t)$ are written as,
\begin{equation}
\label{eq:QH_0deg}
A_{QH}^{0^\circ}(t) = \frac{A_{ox}}{\sqrt{2}}\exp^{i\delta_x}\exp^{i\omega t} + \frac{A_{oy}}{\sqrt{2}}\exp^{i\delta_y}\exp^{i\omega t} \exp^{i\phi}
\end{equation}
\begin{equation}
\label{eq:QH_90deg}
A_{QH}^{90^\circ}(t) = \frac{A_{ox}}{\sqrt{2}}\exp^{i\delta_x}\exp^{i\omega t}\exp^{i\phi} + \frac{A_{oy}}{\sqrt{2}}\exp^{i\delta_y}\exp^{i\omega t} 
\end{equation}
Here $\phi$ represents the phase difference between the in-phase and quadrature signals (which is $90^\circ$) of the input signals that are fed to the QH. Table \ref{tab:QH_op} shows the expected values at the output of QH when linearly polarized (LP) and circularly polarized (CP) signal are received by the CLPDA.
\begin{table}[htbp]
\caption{Output of the quadrature hybrid when linear and circular polarized signals are fed to it}
\begin{center}
\begin{tabular}{c l c c}
\hline
\\
\textbf{S. No.} & \textbf{Tx Signal} & \textbf{$A_{QH}^{0^\circ}(t)$} & \textbf{$A_{QH}^{90^\circ}(t)$} \\ \hline
\\
1. & LP (horizontal) & $\frac{A_o}{\sqrt{2}}$ & $\frac{A_o}{\sqrt{2}}$ \\  
2. & LP (vertical) & $\frac{A_o}{\sqrt{2}}$ & $\frac{A_o}{\sqrt{2}}$ \\  
3. & CP (left) & $\frac{A_o}{\sqrt{2}}$ & $0$ \\  
4. & CP (right) & $0$ & $\frac{A_o}{\sqrt{2}}$  \\
\\
\hline
\end{tabular}
\end{center}
\label{tab:QH_op}
\end{table}

Since the orientations of the two arms of a CLPDA are orthogonal to each other, at any instant, the signal received by one arm will have a phase difference of $90^\circ$ with respect to the other. When two such signals with a $90^\circ$ phase difference are fed to a QH, then one of its outputs would be maximum while the other would be minimum depending upon the nature of circular polarization of the wave (Table \ref{tab:QH_op}).

\subsection{Error in DCP estimation}
\label{subsec:error_DCP}
\begin{figure}[!t]
\centering
\centerline{\includegraphics[width=0.8\textwidth]{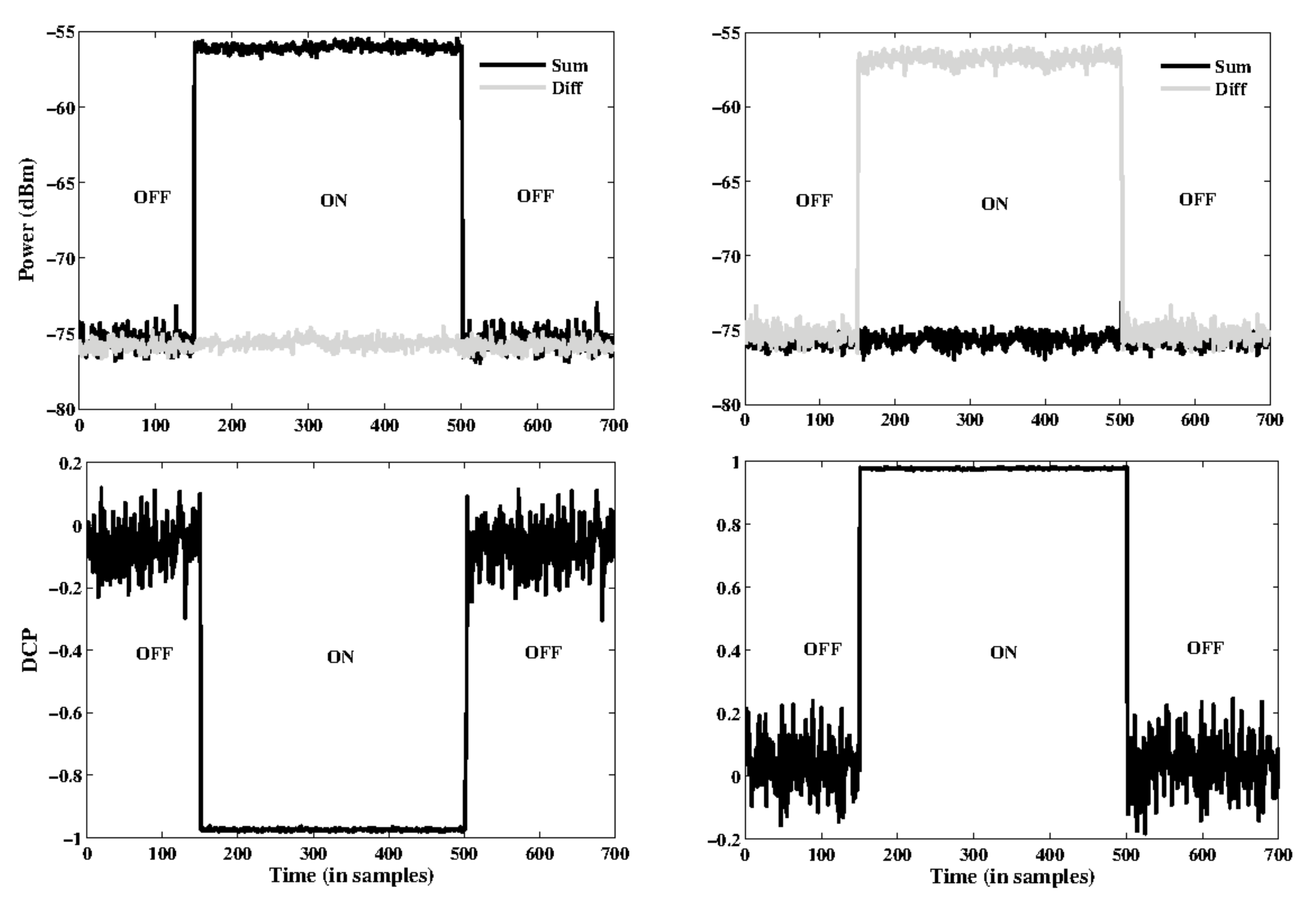}}
\caption{Upper panel: Power received from the two output ports of QH, $P_{QH}^{0^\circ}(t)$ (`black') and $P_{QH}^{90^\circ}(t)$ (`gray'), when CP signal was received by the CLPDA. The left and right panels correspond to left and right CP signals, respectively. Lower panel: Estimated DCP for the same.}
\label{fig:DCP_estimate_CP}
\end{figure}

\begin{figure}[!t]
\centering
\centerline{\includegraphics[width=0.7\textwidth]{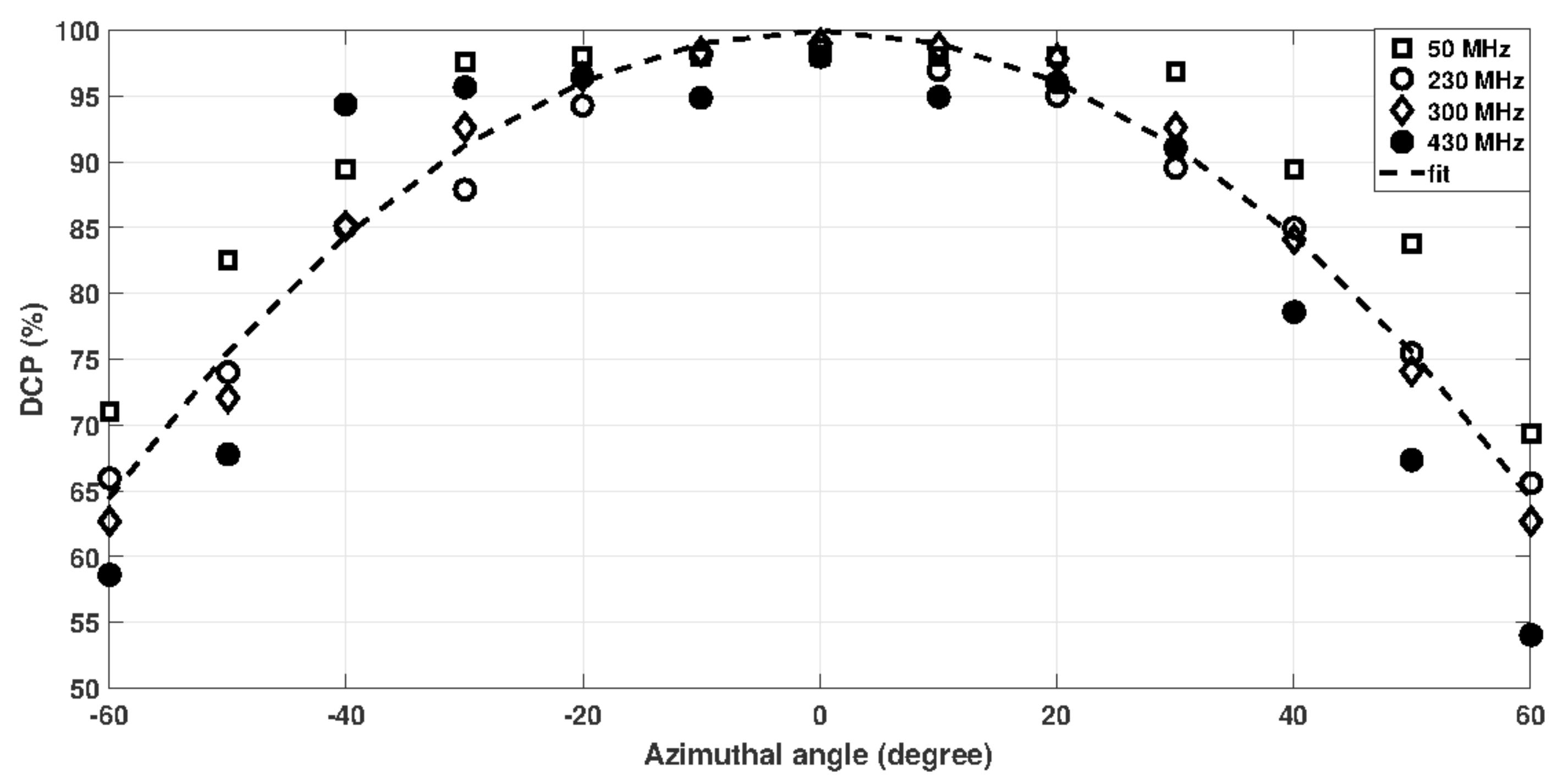}}
\caption{Measured DCP as a function of azimuthal angle. The open square, circle, diamond and filled circle represents the measurements taken at 50, 230, 300 and 430 MHz, respectively. The dashed line is the custom fit to the average DCP values; equation of the form $a cos^b \theta + c$ was used to obtain it. The goodness of fit ($\chi^2$) was maximum ($\approx 0.98$) when the parameters $a$, $b$, and $c$ were equal to 98, 0.65, and 1.9, respectively.}
\label{fig:DCP_Cosfit}
\end{figure}

In 1852, G. Stokes \cite{Collett92} 
proved that the polarization behavior of an EM wave can be represented by four parameters, viz. $I, Q, U,$ and $ V$; they are called as the Stokes parameters. Eqn. \ref{eq:Stokeseqn} holds good for any state of polarization of an EM wave.
\begin{equation}
\label{eq:Stokeseqn}
I^2 \geq Q^2+U^2+V^2
\end{equation}
Here $I$ is the total intensity, $Q$ is the intensity of linear polarization in horizontal or vertical orientation, $U$ is the intensity of linear polarization in $\pm 45^\circ$ and $V$ is the intensity of circular polarization present in the signal. The degree of circular polarization (DCP) is defined as,
\begin{equation}
\label{eq:dcp}
DCP=\frac{|V|}{|I|}
\end{equation}
The values of DCP range between 0 and 1, where 0 represents a wave in unpolarized state and 1 represents it in completely circularly polarized state. To estimate accurately the Stokes parameters and the DCP, the contributions of the antenna and the backend electronics to the circular polarization (CP) have to be determined precisely. For the latter, broadband helical antennas that were available in the observatory were used to generate left and right hand CP signals. The percentage of CP was measured by transmitting a CP wave using the above  helical antennas and measured the output levels of the horizontal and vertical arms of the CLPDA. The DCP values measured with LCP and RCP helical antenna were $\approx$ 100\%, and 98\% circularly polarized, respectively at 430 MHz. Fig. \ref{fig:DCP_estimate_CP} shows the sample test result obtained for $0^\circ$ azimuthal angle. The positive and negative DCP values correspond to LCP and RCP, respectively.
The test was carried out by varying the azimuthal angle for different frequencies within the OB. Fig. \ref{fig:DCP_Cosfit} shows the measured DCP values along with the mean fit. The latter varies from 100\% to 80\% over an azimuthal angle range of $\pm 45^\circ$ with respect to the reference azimuthal angle ($0^\circ$). 

\begin{figure}[!t]
\centering
\centerline{\includegraphics[width=0.7\textwidth]{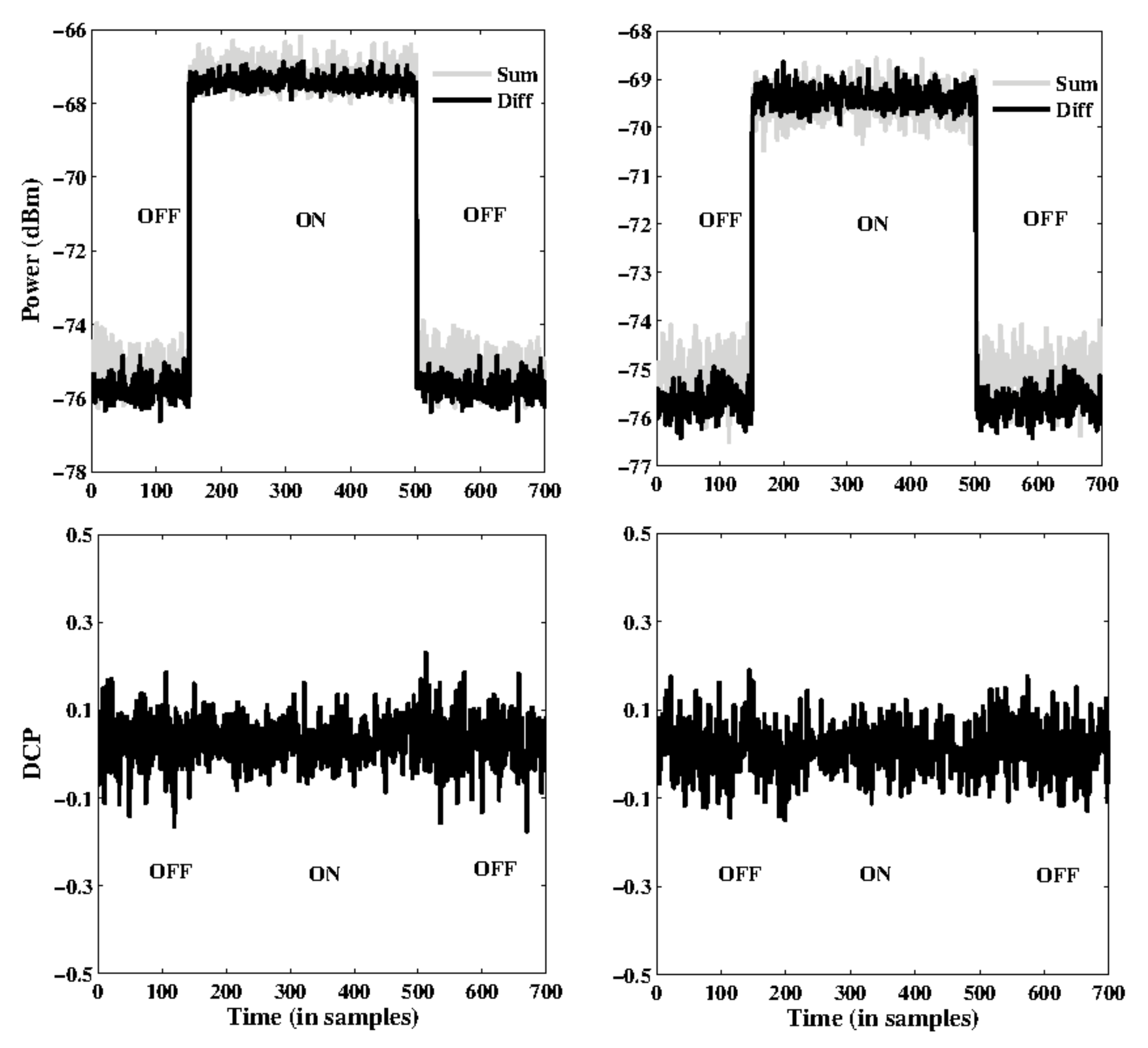}}
\caption{Upper panel: Power received by two outputs of QH, $P_{QH}^{0^\circ}(t)$ (`black') and $P_{QH}^{90^\circ}(t)$ (`gray') when LP signal was received with CLPDA. The left and right panels correspond to the horizontal and vertical polarized signals, respectively. Lower panel: Estimated DCP.}
\label{fig:DCP_estimate_LP}
\end{figure}

Since our aim is to detect CP signal with the above system, linearly polarized (LP) signals were also transmitted using broadband LP antennas and were received with the CLPDA to measure the level of LP contribution to CP. The values obtained at 430 MHz are shown in Fig. \ref{fig:DCP_estimate_LP}. The estimated rms error was $\approx 3\%$. The latter was independently verified with the Galactic Center observations as well.
For calibration, the median subtraction technique was used, i.e., the data obtained immediately after the commencement (and free from interference) of the observation was subtracted from the data obtained throughout the observation from each signal path. This was done to compensate for the losses and to nullify the frequency characteristics of the electronics, cables, etc.

\begin{figure}[!t]
\centering
\centerline{\includegraphics[width=0.7\textwidth]{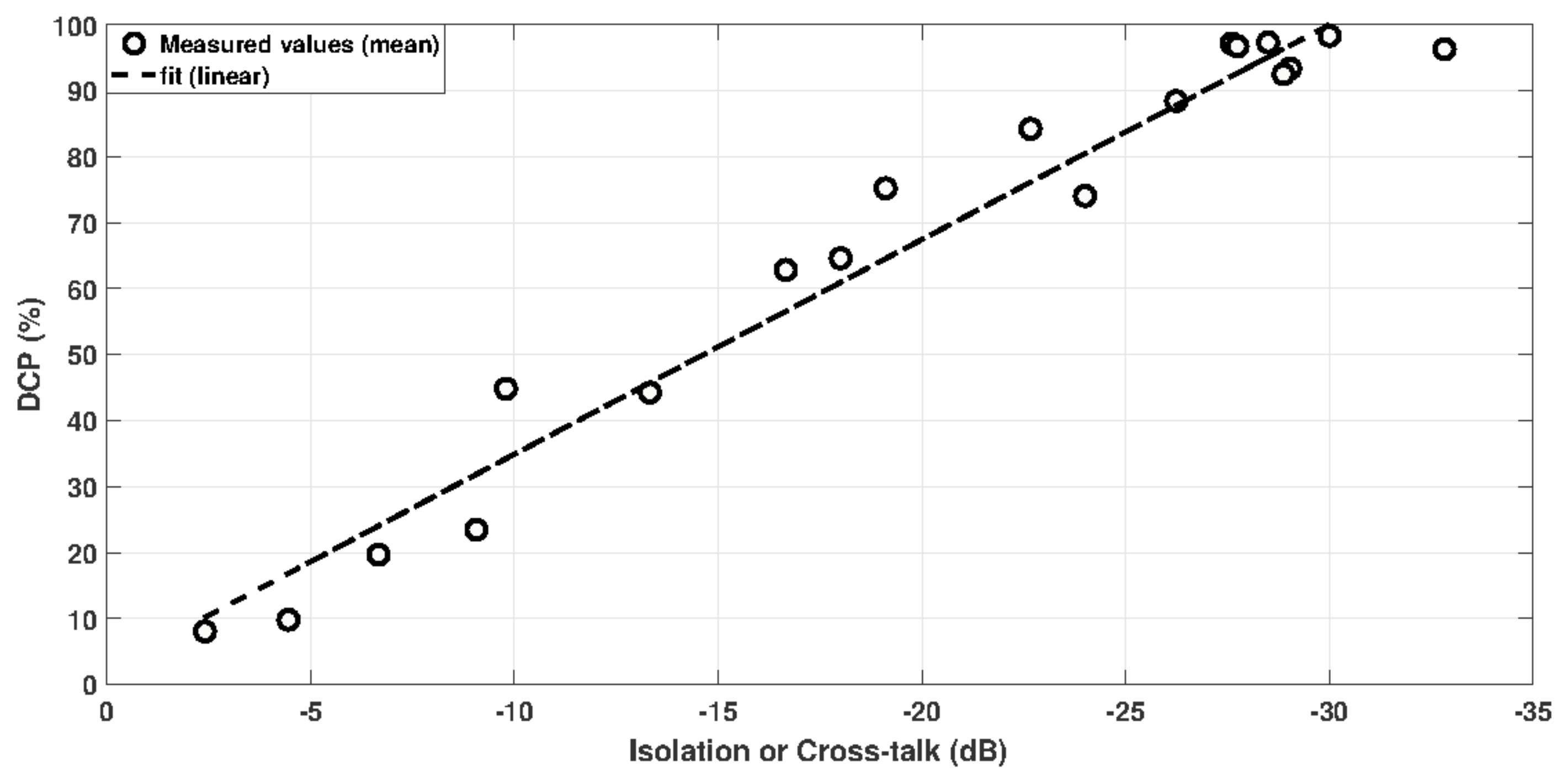}}
\caption{DCP versus Cross-talk. The dashed line is the linear fit ($y = mx +c$) to the mean DCP values. The best fit ($\chi^2 \approx 0.96$) gave -3.3 and 2.0 for m and c, respectively.}
\label{fig:IsoVsDCP}
\end{figure}

It should also be noted here that any linearly polarized component (Stokes-Q and U) would vanish in the intended OB (i.e. meter wavelengths) because of the Faraday rotation (\cite{Hatanaka1956,Grognard1973}); and therefore the emission in total intensity (Stokes-I) and circular polarized intensity (Stokes-V) can only be observed. Therefore, careful attention was given to DCP measurements with CP signals.  Further we plotted the DCP values against the polarization cross-talk making use of their azimuthal angular distribution; Fig. \ref{fig:IsoVsDCP} shows the linear fit between DCP and cross-talk. Since a conservative approach is useful generally, the cross-talk values of the horizontally oriented transmitter configuration was considered for the fit. The plot indicates that the higher DCP can be measured with better accuracy since the corresponding cross-talk is lower; for instance, -30 dB cross-talk corresponds to 0.2 \% error in the mean DCP estimate.

\section{Trial Observations}
\label{sec:obs}
\begin{figure}[!t]
\centering
\centerline{\includegraphics[width=0.7\textwidth]{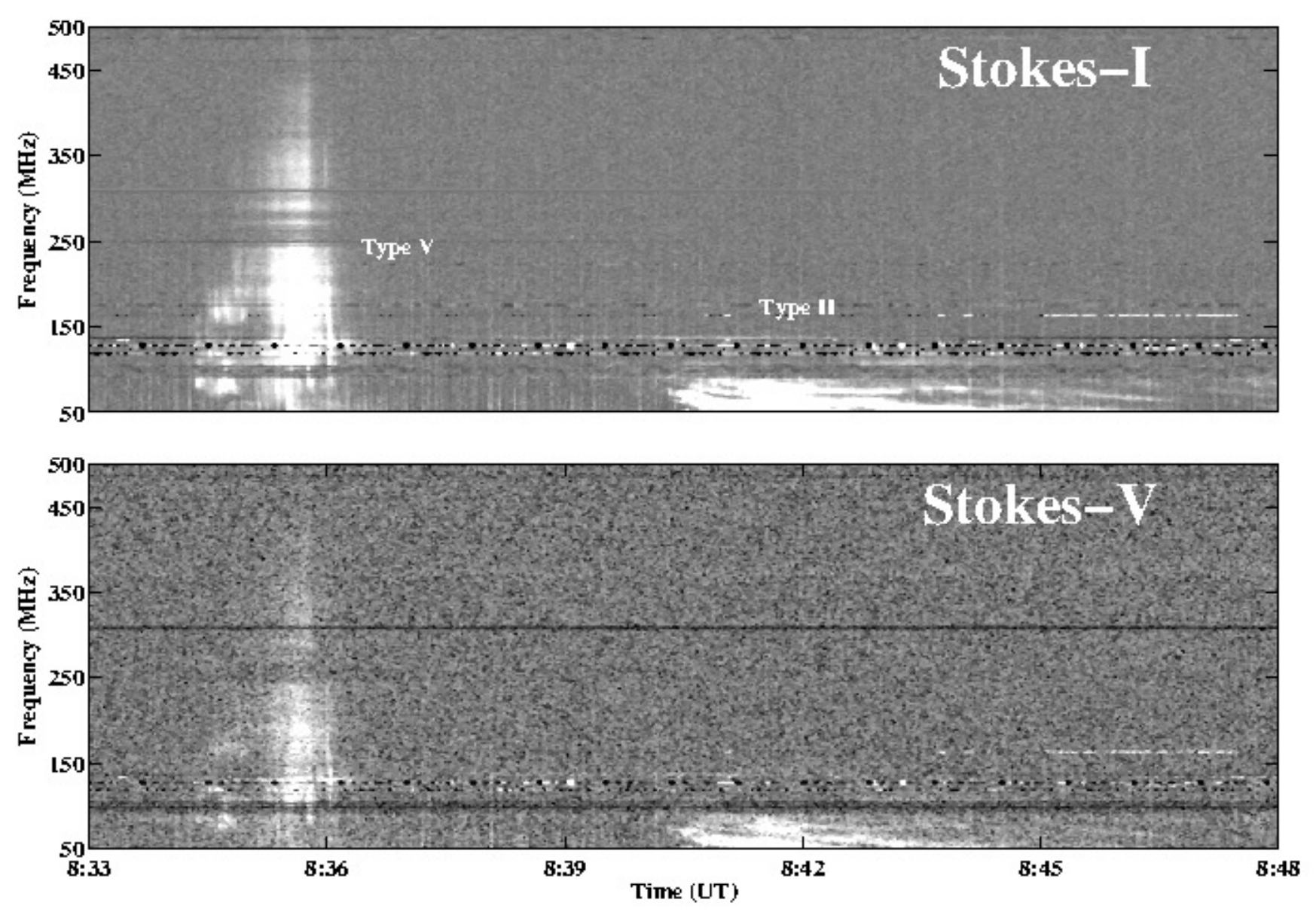}}
\caption{Stokes-I and Stokes-V spectra of a type-V and a type-II burst observed on May 02, 2016. The horizontal lines seen at around 100 MHz and 310 MHz are due to local RFI and are excised.}
\label{fig:typeV_II_StI_StV}
\end{figure}

In order to examine the observing capability, after its construction in mid 2016, the SP set-up was kept for observations at the Gauribidanur observatory, with an intention to determine the Signal-to-Noise Ratio (SNR) and Dynamic Range (DR) of the system. The instrument observed an event on May 02, 2016; Fig. \ref{fig:typeV_II_StI_StV} shows the Stokes-I and Stokes-V spectra of that.
The radio enhancement seen during 08:34 - 08:36 UT is a type-V radio burst\footnote{\url{ftp://ftp.swpc.noaa.gov/pub/indices/events/20160502events.txt}.}; this type is due to synchrotron radiation from the non-thermal electrons that travel at near relativistic speeds \cite{stewart1965solar} and spiral around the curved magnetic field lines in the solar atmosphere. The radio enhancement seen during 8:41 - 8:45 UT is a type-II burst; this type is due to shocks generated by CMEs \cite{Gopal2006}.

\begin{figure}[!t]
\centering
\centerline{\includegraphics[width=0.6\textwidth]{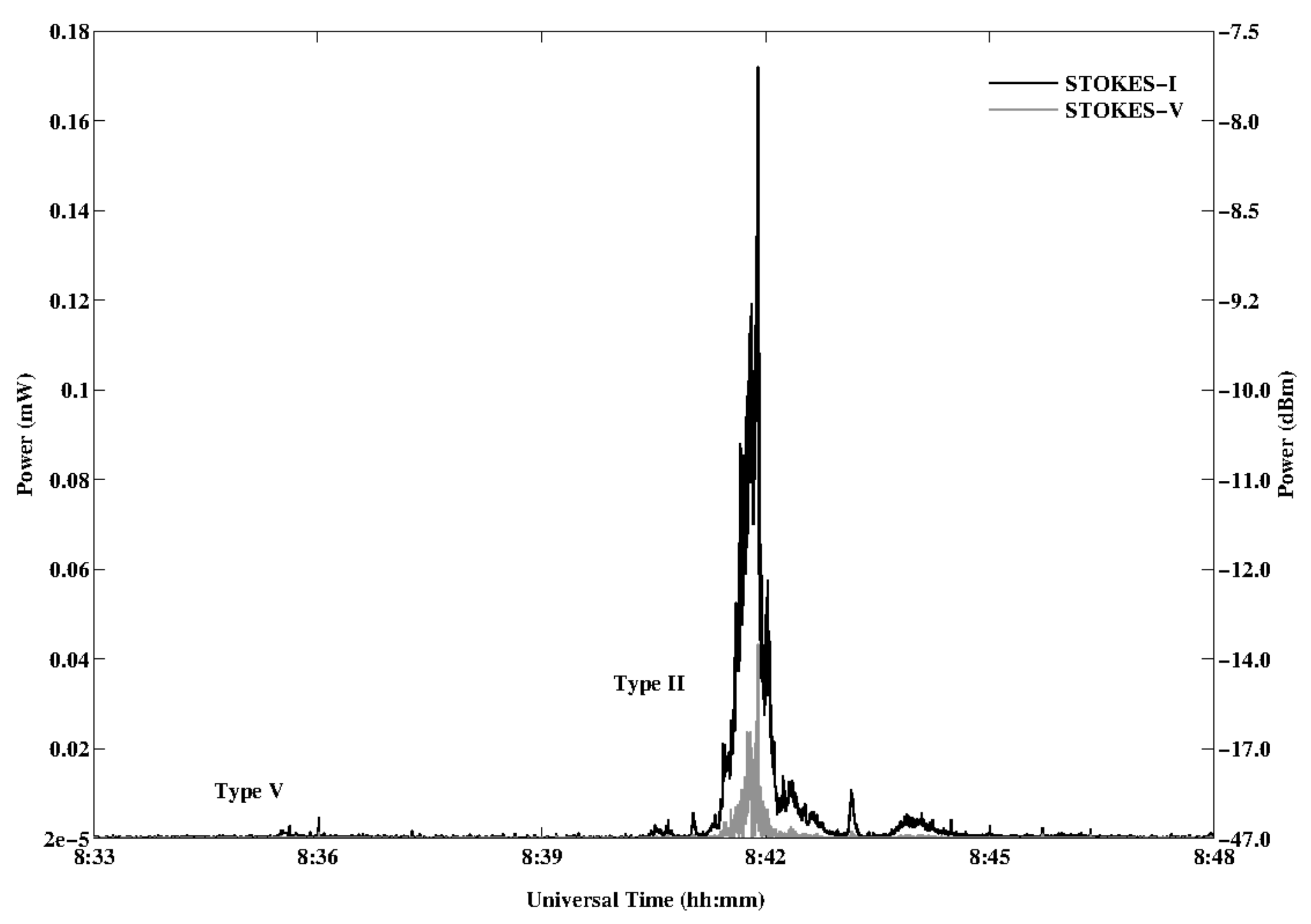}}
\caption{Light curves (Stokes-I and Stokes-V superimposed) of Fig. \ref{fig:typeV_II_StI_StV} at 50 MHz.}
\label{fig:typeV_II_light_curve}
\end{figure}

In order to determine the DR from the spectra, both Stokes-I and Stokes-V light curves corresponding to 50 MHz were plotted; the result is shown in Fig. \ref{fig:typeV_II_light_curve}.
From the Stokes-I profile of type-II, it can be inferred that the SP has a DR of about 40 dB. The SNR calculated from the type-II was found to vary between $\approx$ 30 dB (at 50 MHz) and 5 dB (at 100 MHz). Besides, the Stokes-I spectrum of type-V is a vivid example for the frequency response of the SP : It works well in the 50 - 500 MHz range. The spectra and the light-curves were obtained from the recorded signal using the following set of equations:
\begin{equation}
\label{eq:P_QH_0deg}
P_{QH}^{0^\circ}(t){(in\,\,mW)}=10^{\frac{P_{QH}^{0^\circ}(t){(in\,\,dBm)}}{10}}
\end{equation}
\begin{equation}
\label{eq:P_QH_90deg}
P_{QH}^{90^\circ}(t){(in\,\,mW)}=10^{\frac{P_{QH}^{90^\circ}(t){(in\,\,dBm)}}{10}}
\end{equation}
\begin{equation}
\label{eq:StokesI}
I{(in\,\,mW)}= \Bigl| \frac{P_{QH}^{0^\circ}(t){(in\,\,mW)}+P_{QH}^{90^\circ}(t){(in\,\,mW)}}{2} \Bigl|
\end{equation}
\begin{equation}
\label{eq:StokesV}
V{(in\,\,mW)}=\Bigl| \frac{P_{QH}^{0^\circ}(t){(in\,\,mW)}-P_{QH}^{90^\circ}(t){(in\,\,mW)}}{2} \Bigl|
\end{equation}
\begin{equation}
\label{eq:IdB_ImW}
I(in\,\,dBm)=10\,log\,I(in\,\,mW)
\end{equation}
\begin{equation}
\label{eq:VdB_VmW}
V(in\,\,dBm)=10\,log\,V(in\,\,mW)
\end{equation}

\subsection{Estimation of Coronal Magnetic Field Strength}
\label{subsec:mag_field}
Since the SP was perceived to have a reasonably good DR and SNR, it has been put into regular observing schedule at the Gauribidanur observatory. Finally, to demonstrate the usability of the observed spectral data, one of their invaluable applications in the field of solar radio astronomy, viz., the estimation of coronal magnetic field strength (B), is exemplified below.

Theoretical studies suggest that B can be estimated from the observed DCP of a radio burst (ex: type-III, type-V) using Eqn. \ref{eq:BvsDCP}, if (i) the burst is due to second harmonic plasma emission \cite{mclean1985solar} and (ii) its associated source region lies close to the solar limb;
\begin{equation}
\label{eq:BvsDCP}
B=\frac{f_p \times {DCP}}{2.8 \times a(\theta)}~~ (Gauss)
\end{equation}
where, $f_{p}$ is the plasma frequency and $\theta$ is the viewing angle of the observer with respect to the radial direction of the magnetic field. The function $a(\theta) \approx 1$ for sources located near the solar limb.

\begin{figure}[!t]
\centering
\centerline{\includegraphics[width=0.6\textwidth]{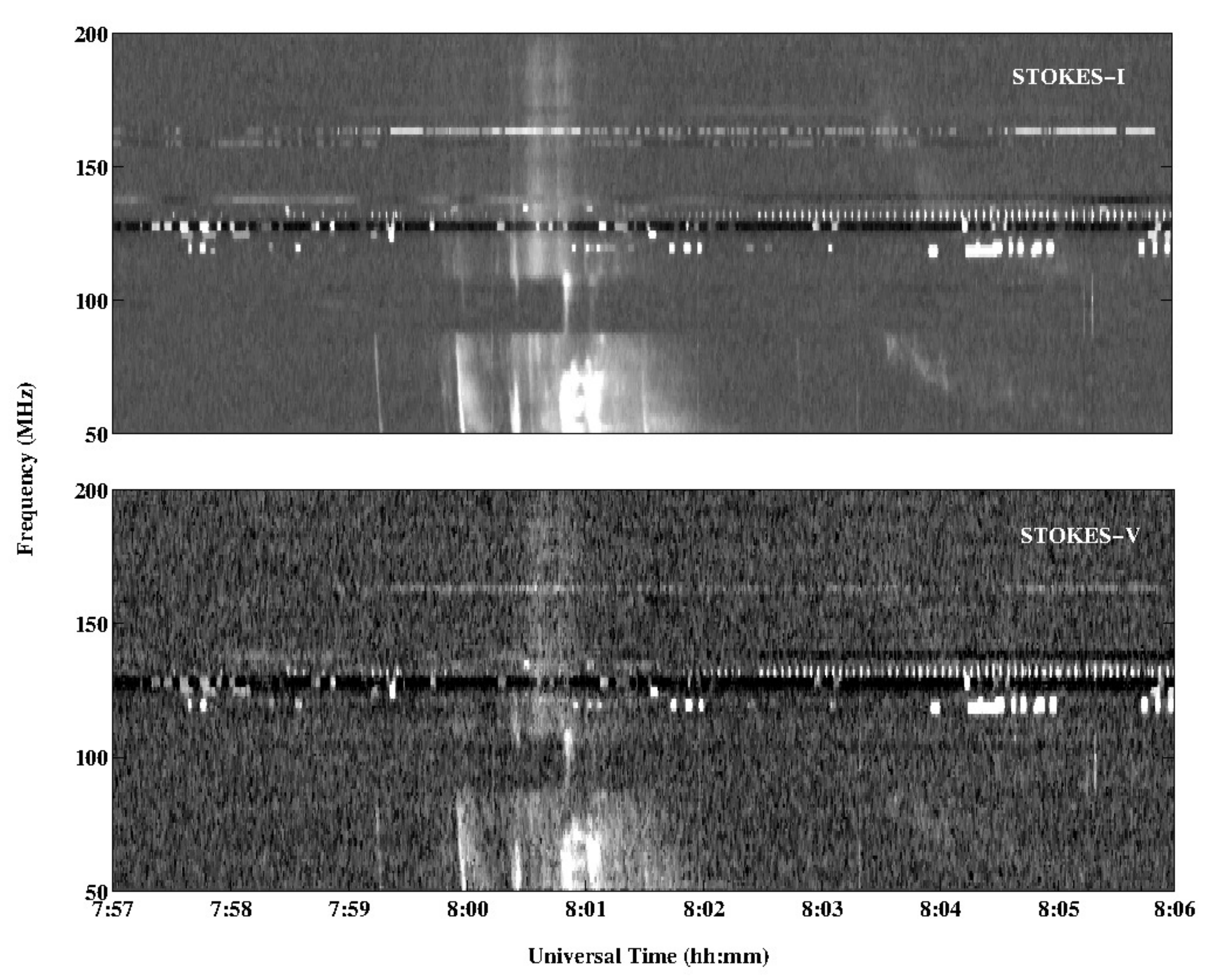}}
\caption{Stokes-I and Stokes-V spectra of a type-V and a type-II burst observed on March 30, 2018.}
\label{fig:spec_type-V}
\end{figure}

\begin{figure}[!t]
\centering
\centerline{\includegraphics[width=0.5\textwidth]{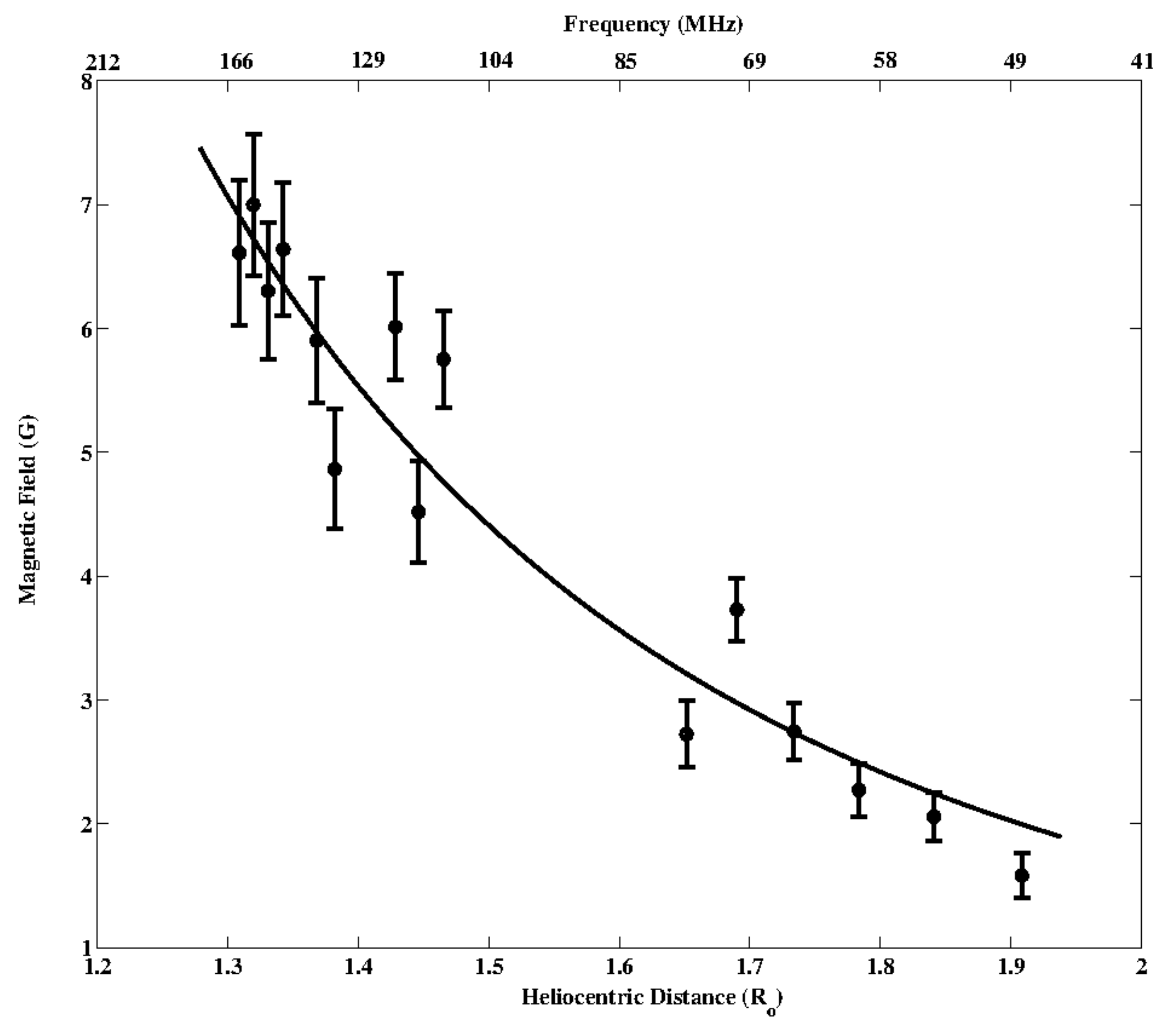}}
\caption{Magnetic field strength (y-axis) associated with a type-V radio burst, as a function of heliocentric distance (bottom x-axis) / observing frequency (top x-axis).}
\label{fig:type-V_B}
\end{figure}

We searched for an event in the spectral database and found that the type-V radio burst observed with the SP on March 30, 2018 satisfies the above criteria. The spectra are shown in Fig. \ref{fig:spec_type-V}. The radio enhancement seen during 8:00 - 8:02 UT is the type-V burst. This was associated with C4.6 class flare erupted from the active region 12703 located at S10E70. 
Using Eqn. \ref{eq:dcp} (section \ref{subsec:error_DCP}), the DCP of the  above type-V burst was calculated at discrete observing frequencies and plotted against the heliocentric distance ($r$) using Newkirk's density model \cite{newkirk1961solar}; this gives $r$ values in the range 1.3 - 1.9 $R_{\odot}$ (where $1\,R_\odot=6.96\times 10^5~km=$ radius of solar photosphere). 
\begin{figure}[!t]
\centering
\centerline{\includegraphics[width=0.6\textwidth]{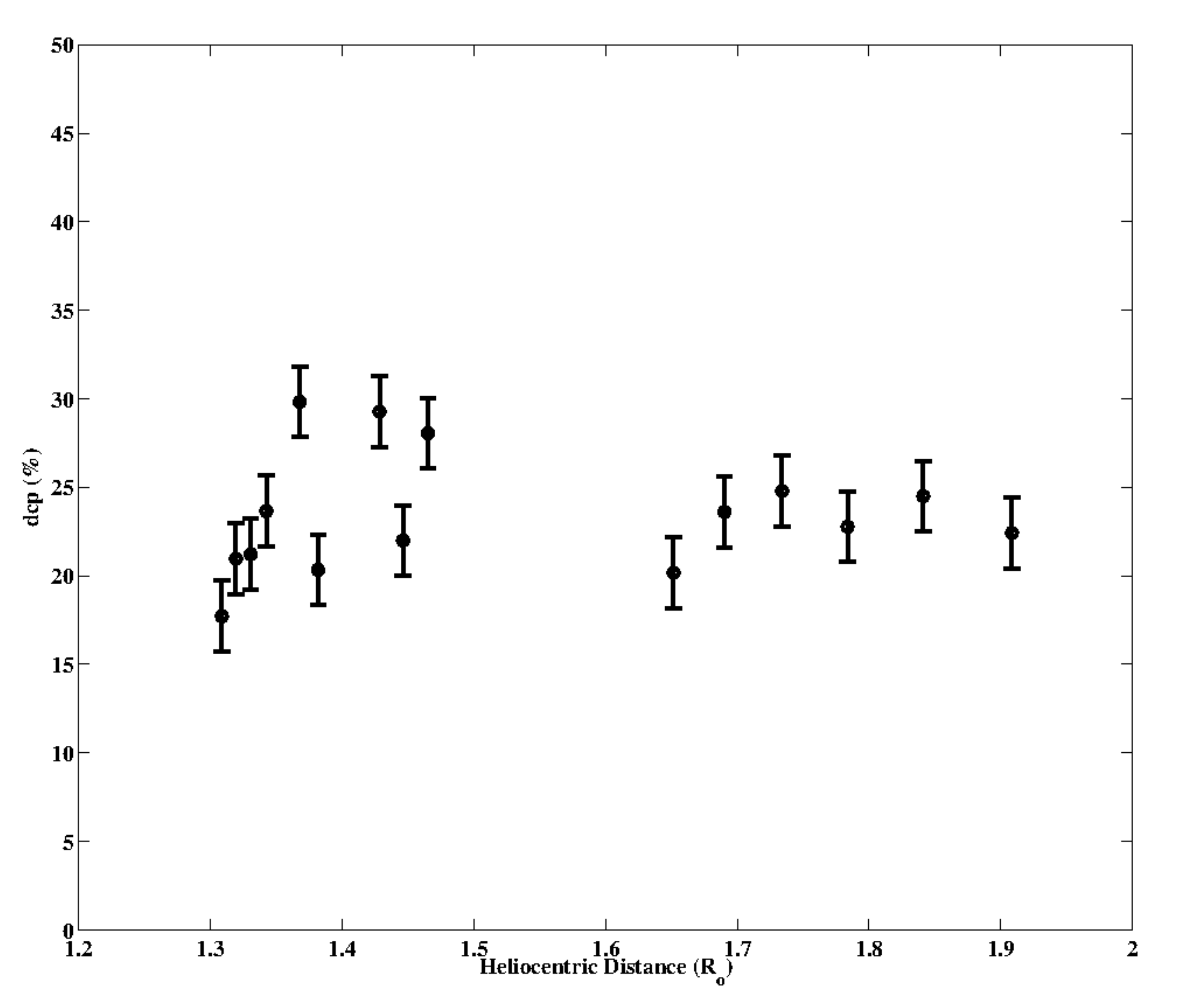}}
\caption{DCP (with an uncertainty of 3\%) as a function of heliocentric distance, of the type-V burst shown in Fig. \ref{fig:spec_type-V}. The data points between 1.46~$R_\odot$ (108 MHz) and 1.64 $R_\odot$ (80 MHz) were omitted as they were influenced by RFI. }
\label{fig:DCP_typeV}
\end{figure}
The result (Fig. \ref{fig:DCP_typeV}) shows that the DCP varies between 17~\% and 30 \% with mean around 24 \%. Using the latter and Eqn. \ref{eq:BvsDCP}, B was estimated; Fig. \ref{fig:type-V_B} shows its distribution as a function of $r$.
The power-law fit to the data points gives $B(r)=16.8 \pm 0.5\,r^{-3.3}$ Gauss. This distribution is similar to those published earlier : for example, Patzold et al. \cite{Patzold1987} found that $B(r)$ varies as $r^{-2.7}$ in the middle corona; Similarly, Lin et. al, \cite{lin2004coronal} had estimated a magnetic field strength of about 4 G at 1.2 $R_{\odot}$. We would like to note here that the type-V radio burst observed on May 03, 2016 (Fig. \ref{fig:typeV_II_StI_StV}) could not be used to determine B with the help of Eqn. \ref{eq:BvsDCP} because its source region was located at the central portion of the Sun's disk on that day.

\section{Discussions \& Conclusions}
\label{sec:vista}
We designed and fabricated a CLPDA that works in the 50 - 500 MHz frequency range; the design constraints, the procedure to tune its impedance and to minimize its dimension, are explained. Throughout the OB, the CLPDA has a directive gain of about 6.6 dBi, return loss $\lesssim$ -10 dB, and a polarization leakage or cross-talk $\lesssim$ -27 dB at the reference position angle (i.e. azimuthal angle $= 0^\circ$). The latter is about 10 dB lesser than the commercially available ones; this is most likely due to the usage of rectangular bars as transmission lines in stead of the generally used square tubes to fabricate the LPDA / CLPDA. The variation of the cross-talk as a function of azimuthal angle ($\theta$) was also measured for the transmitter kept in both horizontal and vertical positions; the custom fit to the mean cross-talk values follow the form $cos^{b}\theta$, where $b$ is equal to 0.65 and 0.85, respectively. A spectro-polarimeter was set up using the CLPDA, an analog receiver and a digital receiver (Spectrum Analyzer), and the system as a whole was characterized; the analog receiver has a noise figure of $\approx 3$ dB and a $T_{rcvr}$ of about 290 K. The digital receiver has an instantaneous bandwidth of $\approx$ 1.1 MHz. The receiver was found to have a sensitivity of $\approx$ 0.53 sfu at 50 MHz, and 53 sfu at 500 MHz, for the parameters mentioned above and for an effective integration time of 100 $\mu s$. Then, its polarization detection capability was studied : The CP wave was transmitted using RCP and LCP antennas and the same were measured using the CLPDA as a function of azimuthal angle. The mean fit to the average DCP detected was plotted and was found to follow the form $cos^b\theta$, where $b$ is equal to 0.65; the magnitude was found to vary between 100\% and 80\% over an azimuthal angle of $\pm 45^\circ$ with respect to the reference azimuthal angle ($0^\circ$). Further the DCP values were plotted against the polarization cross-talk making use of their azimuthal angular distribution; the fit was found to be linear (slope = -3.5). To have a conservative approach, the cross-talk values of the horizontally oriented transmitter configuration was considered for the fit. It was inferred from the plot that lower cross-talk values can give rise to better accuracy in the estimated DCP. After characterization, trial observations of the Sun were carried out; using the type-II spectra observed on May 05, 2016, the SNR and dynamic range were determined and the values are about 30 dB and 40 dB, respectively, at 50 MHz. To demonstrate the instrument capability, the Stokes-I and Stokes-V spectrum of the type-V burst obtained on March 30, 2018 with the SP was used to calculate the DCP; the uncertainty in the measurement is $\approx 3\%$. Using the DCP, B (as a function of heliocentric height) associated with the type-V emission was determined; the distribution ($B(r)=16.8\pm 0.5 \,r^{-3.3}$ G) is in good agreement with those reported earlier. 

Measuring DCP using the co-located linearly polarized feeds of the CLPDA is always advantageous because each one of them can be oriented along the major and minor axes of the polarization ellipse; the system then responds only to the circular polarization \cite{Morris1964} which would be more suitable for observing the solar radio bursts as compared to observing with circularly polarized feeds. In such a case, the system shall respond as a zero-baseline correlation interferometer \cite{Morris1964} with two arms of the CLPDA as two feeds. Otherwise, one would end up with a modulated output profile from the interferometer, if the orthogonal feeds are kept apart. Again, if an array of CLPDAs is built with tracking facility, then the DCP can be measured with minimum error as the source will always be on the maximum response portion of the CLPDA; additionally an array always gives better sensitivity, gain, etc. and therefore, the weak non-thermal energy releases that are expected to prevail the corona all the time, can be observed. A new FPGA based digital backend receiver is also being planned at the Gauribidanur observatory to obtain the data with better spectral and temporal resolution, sensitivity, signal-to-noise ratio, dynamic range, etc. as well. Since the CLPD design discussed in this article has less polarization cross-talk, less design cost and does not require BALUN, etc., it may be utilized for the proposed / upcoming radio array facilities (\cite{delera2015,Ivanov2019}). Methods to lower the cross-talk further and to have a uniform value over the entire HPBW may be evolved to reduce the error in DCP estimates. Also, it would be a very good advancement in technology, if an optimal design is conceived in which the CLPDA can be used as a feed in dish antennas.

\section*{Acknowledgment}

We would like to thank the staff of the Gauribidanur observatory for their continuous support in fabricating the antennas, carrying out various tests, building various modules, and for their help in maintaining the antenna, receiver systems, and in carrying out the observations regularly. 
The authors thank Prof. R. Ramesh for his insightful comments during the development of the antenna. AK acknowledges K. S. Raja, P. Kishore and K. Hariharan for discussions; and also the European Research Council (ERC) functioning under the European Union’s Horizon 2020 Research and Innovation Programme Project SolMAG 724391. CK expresses his sincere thanks to Prof. Satyanarayanan for having proofred the manuscript.

\bibliographystyle{unsrt}  


\end{document}